\documentclass{llncs}
 
\usepackage{microtype}
\usepackage{url}
\usepackage[breaklinks]{hyperref}   

\usepackage[utf8]{inputenc}
\usepackage{amsmath}
\usepackage{txfonts}
\usepackage{cancel}
\usepackage{tikz}
\usepackage{stmaryrd}
\usepackage{etoolbox}
\usepackage{float}
\usepackage{complexity}
\usepackage{thmtools}
\usepackage[numbers]{natbib}
\usepackage[ligature,reserved,inference]{semantic}
\usepackage{at}
\usetikzlibrary{positioning}
\usetikzlibrary{automata}

\newtoggle{fullversion}
\toggletrue{fullversion}

\title{On Reducing Linearizability to State Reachability\thanks{This work is
supported in part by the VECOLIB project (ANR-14-CE28-0018).}}

\author{
  Ahmed Bouajjani\inst{1}
  \and Michael Emmi\inst{2}
  \and Constantin Enea\inst{1}
  \and Jad Hamza\inst{1}
}
\institute{
  LIAFA, Université Paris Diderot, France
  \and IMDEA Software Institute, Spain
}

\makeatletter
\renewcommand\bibsection%
{
  \section*{\refname
    \@mkboth{\MakeUppercase{\refname}}{\MakeUppercase{\refname}}}
}
\makeatother

\newcommand{\set}[1]{\{{#1}\}}
\newcommand{\sect}[1]{Section~\ref{#1}}
\newcommand{\lem}[1]{Lemma~\ref{#1}}
\newcommand{\coro}[1]{Corollary~\ref{#1}}
\newcommand{\thm}[1]{Theorem~\ref{#1}}

\newcommand{\modulo}[2]{{#1}\ mod\ {#2}}
\newcommand{\fig}[1]{Fig.~\ref{#1}}
\newcommand{\ie}{i.e.,\ }
\newcommand{\resp}{resp.,\ }

\newcommand{\domain}{\mathbb{D}}

\newcommand{\ds}{\mathcal{S}}
\newcommand{\queue}{{\sf Queue}}
\newcommand{\stack}{{\sf Stack}}
\newcommand{\mutex}{{\sf Mutex}}

\newcommand{\register}{{\sf Register}}

\newcommand{\meth}{m}
\newcommand{\Meth}{\mathbb{M}}
\newcommand{\meths}{M}

\newcommand{\emptyseq}{\epsilon}
\newcommand{\useq}{u}
\newcommand{\vseq}{v}
\newcommand{\wseq}{w}
\newcommand{\entails}{\Rightarrow}
\renewcommand{\cc}{\cdot}
\newcommand{\conditions}{Guard}
\newcommand{\constructor}{Expr}

\newcommand{\matchset}[1]{M{#1}}
\newcommand{\notafter}{\not >}

\newcommand{\dw}[1]{\llbracket {#1} \rrbracket}

\newcommand{\fresh}{x}
\newcommand{\dat}{d}
\newcommand{\dats}{D}
\newcommand{\domof}[1]{\domain_{#1}}

\newcommand{\rul}{R}
\newcommand{\ruler}{rule}
\newcommand{\getrule}{{\tt last}}
\newcommand{\applyrule}[1]{\xrightarrow{#1}}

\newcommand{\distinguished}{differentiated}
\newcommand{\execution}{execution}
\newcommand{\history}{history}
\newcommand{\sbs}{step-by-step linearizable}

\newcommand{\wellformed}{non-ambiguous}

\newcommand{\wellformedness}{non-ambiguity}
\newcommand{\tractable}{co-regular}
\newcommand{\dataindependent}{data-independent}
\newcommand{\corresponding}{corresponding}
\newcommand{\methodevent}{method event}

\newcommand{\closedbyerasure}{closed under projection}
\newcommand{\matchingset}{matching set}
\newcommand{\dataword}{sequential execution}

\newcommand{\datastructure}{data structure}
\newcommand{\elastic}{closed under projection}
\newcommand{\gap}{gap}
\newcommand{\leftrightconstraint}{left-right constraint}

\newcommand{\implementation}{implementation}

\newcommand{\impl}{\mathcal{I}}

\newcommand{\oenqueue}{Enq}
\newcommand{\enqueue}{Enq}
\newcommand{\dequeue}{Deq}
\newcommand{\deqempty}{DeqEmpty}
\newcommand{\rstart}{R_0}
\newcommand{\renq}{R_{\oenqueue}}
\newcommand{\renqdeq}{R_{EnqDeq}}
\newcommand{\rempty}{R_{DeqEmpty}}

\newcommand{\push}{Push}
\newcommand{\opush}{Push}
\newcommand{\pop}{Pop}
\newcommand{\popempty}{PopEmpty}
\newcommand{\rpush}{R_{\opush}}
\newcommand{\rpushpop}{R_{PushPop}}
\newcommand{\rpopempty}{R_{PopEmpty}}

\newcommand{\writevar}{Write}
\newcommand{\readvar}{Read}
\newcommand{\rreg}{R_{WR}}

\newcommand{\lock}{Lock}
\newcommand{\olock}{Lock}
\newcommand{\unlock}{Unlock}
\newcommand{\rlock}{R_{\olock}}
\newcommand{\rlockunlock}{R_{LU}}

\newcommand{\projections}[1]{{\sf proj}({#1})}
\newcommand{\removefrom}[2]{{#1} \smallsetminus {#2}}

\newcommand{\linable}[2]{{#1} \sqsubseteq {#2}}
\newcommand{\notlinable}[2]{{#1} \not\sqsubseteq {#2}}
\newcommand{\op}{o}
\newcommand{\ops}{O}

\newcommand{\winning}{\ {\bf W}\ }
\newcommand{\notwinning}{\ \cancel{\bf W}\ }
\newcommand{\winningp}[2]{\ {\bf W_{{#1},{#2}}}\ }

\newcommand{\hist}{h}
\newcommand{\hists}{H}
\newcommand{\project}[2]{{#1}_{|#2}}
\newcommand{\happenedbefore}{happens-before}
\newcommand{\hb}{\leq_{hb}}

\newcommand{\exec}{e}

\newcommand{\nrule}{n}
\newcommand{\nseq}{k}
\newcommand{\ndats}{m}
\newcommand{\nfact}{l}

\newcommand{\nops}{s}
\newcommand{\nreads}{s}
\newcommand{\ulength}{n}
\newcommand{\bij}{f}

\newcommand{\expr}{E}
\newcommand{\fact}{a}

\newcommand{\dataletter}[2]{{#1}({#2})}
\newcommand{\thealphabet}[1]{\dataletter{\Meth}{#1}}

\newcommand{\lesser}[3]{
  \textsf{matched}({#2},{#3})
}

\newcommand{\aut}{\mathcal{A}}
\newcommand{\autof}[1]{\aut_{#1}}

\newcommand{\restrict}[1]{{#1}_\neq}
\newcommand{\mainmethod}{\Meth_{in}}
\newcommand{\main}{input}

\newcommand{\callevent}[3]{{\tt call_{#3}\ } {#1}({#2})}
\newcommand{\retevent}[3]{{\tt ret_{#3}\ } {#1}({#2})}
\newcommand{\methevent}[2]{\dataletter{#1}{#2}}
\newcommand{\getlabel}{l}

\newcommand{\Op}{\mathbb{O}}
\newcommand{\operation}{operation}

\newcommand{\thewrite}{a}
\newcommand{\theread}{b}
\newcommand{\theempty}{o}
\newcommand{\notall}[2]{{#1}_{#2}}
\newcommand{\ourclass}{\mathcal{C}}
\newcommand{\rename}[1]{r(#1)}
\newcommand{\dwords}{\ds}
\newcommand{\action}{action}
\newcommand{\coregular}{co-regular}

\iftoggle{fullversion}{
  \pagestyle{plain}
}{}

\begin{document}
 
  \maketitle

  \begin{abstract}

  Efficient implementations of atomic objects such as concurrent stacks and
  queues are especially susceptible to programming errors, and necessitate
  automatic verification. Unfortunately their correctness criteria —
  linearizability with respect to given ADT specifications — are hard to verify.
  Even on classes of implementations where the usual temporal safety properties
  like control-state reachability are decidable, linearizability is undecidable.

  In this work we demonstrate that verifying linearizability for certain
  \emph{fixed} ADT specifications is reducible to control-state reachability,
  despite being harder for \emph{arbitrary} ADTs. We effectuate this reduction
  for several of the most popular atomic objects. This reduction yields the
  first decidability results for verification without bounding the number of
  concurrent threads. Furthermore, it enables the application of existing
  safety-verification tools to linearizability verification.

\end{abstract}
  \section{Introduction}

Efficient implementations of atomic objects such as concurrent queues and
stacks are difficult to get right. Their complexity arises from the conflicting
design requirements of maximizing efficiency/concurrency with preserving the
appearance of atomic behavior. Their correctness is captured by
\emph{observational refinement}, which assures that all behaviors of programs
using these efficient implementations would also be possible were the atomic
reference implementations used instead.
Linearizability~\cite{journals/toplas/HerlihyW90}, being an equivalent
property~\cite{journals/tcs/FilipovicORY10,Bouajjani:2015:TRC:2676726.2677002},
is the predominant proof technique: one shows that each concurrent execution
has a linearization which is a valid sequential execution according to a
specification, given by an abstract data type (ADT) or reference implementation.

Verifying automatically\footnote{Without programmer annotation — see
Section~\ref{sec:related} for further discussion.} that all executions of a
given implementation are linearizable with respect to a given ADT is an
undecidable problem~\cite{conf/esop/BouajjaniEEH13}, even on the typical
classes of implementations for which the usual temporal safety properties are
decidable, e.g.,~on finite-shared-memory programs where each thread is a
finite-state machine. What makes linearization harder than typical temporal
safety properties like control-state reachability is the existential
quantification of a valid linearization per execution.

In this work we demonstrate that verifying linearizability for certain
\emph{fixed} ADTs is reducible to control-state reachability, despite being
harder for \emph{arbitrary} ADTs. We believe that fixing the ADT parameter of
the verification problem is justified, since in practice, there are few ADTs
for which specialized concurrent implementations have been developed. We
provide a methodology for carrying out this reduction, and instantiate it on
four ADTs: the atomic queue, stack, register, and mutex.

Our reduction to control-state reachability holds on any class of
implementations which is closed under intersection with regular
languages\footnote{We consider languages of well-formed method call and return
actions, e.g.,~for which each return has a matching call.} and which is
\emph{data independent} — informally, that implementations can perform only
read and write operations on the data values passed as method arguments. From
the ADT in question, our approach relies on expressing its violations as a
finite union of regular languages.

In our methodology, we express the atomic object specifications using inductive
rules to facilitate the incremental construction of valid executions. For
instance in our atomic queue specification, one rule specifies that a dequeue
operation returning empty can be inserted in any execution, so long as each
preceding enqueue has a corresponding dequeue, also preceding the inserted
empty-dequeue. This form of inductive rule enables a locality to the reasoning
of linearizability violations.

Intuitively, first we prove that a sequential execution is invalid if and only 
if some subsequence could not have been produced by one of the rules. Under
certain conditions this result extends to concurrent executions: an execution
is not linearizable if and only if some projection of its operations cannot be
linearized to a sequence produced by one of the rules. We thus correlate the
finite set of inductive rules with a finite set of classes of non-linearizable
concurrent executions. We then demonstrate that each of these classes of
non-linearizable executions is regular, which characterizes the violations of a
given ADT as a finite union of regular languages. The fact that these classes
of non-linearizable executions can be encoded as regular languages is somewhat
surprising since the number of data values, and thus alphabet symbols, is, a
priori, unbounded. Our encoding thus relies on the aforementioned \emph{data
independence} property.

To complete the reduction to control-state reachability, we show that
linearizability is equivalent to the emptiness of the language intersection
between the implementation and finite union of regular violations. When the
implementation is a finite-shared-memory program with finite-state threads,
this reduces to the coverability problem for Petri nets, which is decidable,
and EXPSPACE-complete.

To summarize, our contributions are:
\begin{itemize}

  \item a generic reduction from linearizability to control-state reachability,
  
  \item its application to the atomic queue, stack, register, and mutex ADTs,

  \item the methodology enabling this reduction, which can be reused on other
  ADTs, and

  \item the first decidability results for linearizability without bounding
  the number of concurrent threads.

\end{itemize}
Besides yielding novel decidability results, our reduction paves the way for
the application of existing safety-verification tools to linearizability
verification.

\sect{sec:prelim} outlines basic definitions. \sect{sec:structures} describes a
methodology for inductive definitions of data structure specifications. In
\sect{sec:reduction} we identify conditions under which linearizability can be
reduced to control-state reachability, and demonstrate that typical atomic
objects satisfy these conditions. Finally, we prove decidability of
linearizability for finite-shared-memory programs with finite-state threads in
\sect{sec:decidability}. Proofs to technical results appear in
\iftoggle{fullversion}{the appendix}{the extended version of this
paper~\citep{DBLP:journals/corr/BouajjaniEEH15}}.

\section{Linearizability}
\label{sec:prelim}

We fix a (possibly infinite) set $\domain$ of \emph{data values}, and a finite
set $\Meth$ of \emph{methods}. We consider that methods have exactly one
argument, or one return value. Return values are transformed into argument
values for uniformity.\footnote{%
  Method return values are guessed nondeterministically, and validated at
  return points. This can be handled using the {\tt assume} statements of
  typical formal specification languages, which only admit executions
  satisfying a given predicate. The argument value for methods without argument
  or return values, or with fixed argument/return values, is ignored.}
In order to differentiate methods taking an argument (e.g., the $\enqueue$
method which inserts a value into a queue) from the other methods, we identify
a subset $\mainmethod \subseteq \Meth$ of \emph{\main} methods which do take
an argument. A \emph{\methodevent} is composed of a method $\meth \in \Meth$
and a data value $\fresh \in \domain$, and is denoted
$\methevent{\meth}{\fresh}$. We define the \emph{concatenation} of method-event
sequences $\useq \cc \vseq$ in the usual way, and $\emptyseq$ denotes the empty
sequence.

\begin{definition}

  A \emph{\dataword} is a sequence of \methodevent{s}, 

\end{definition}

The projection $\project{\useq}{\dats}$ of a \dataword{} $\useq$ to a subset
$\dats \subseteq \domain$ of data values is obtained from $\useq$ by erasing
all \methodevent{s} with a data value not in $\dats$. The set of projections of
$\useq$ is denoted $\projections{\useq}$. We write $\removefrom{\useq}{\fresh}$
for the projection $\project{\useq}{\domain\setminus\set{\fresh}}$.

\begin{example}

  The projection $\removefrom{ \dataletter{\enqueue}{1}
  \dataletter{\enqueue}{2} \dataletter{\dequeue}{1} \dataletter{\enqueue}{3}
  \dataletter{\dequeue}{2} \dataletter{\dequeue}{3} }{1}$ is equal to
  $\dataletter{\enqueue}{2} \dataletter{\enqueue}{3} \dataletter{\dequeue}{2}
  \dataletter{\dequeue}{3}$.

\end{example}

We also fix an arbitrary infinite set $\Op$ of operation (identifiers).
A \emph{call \action} is composed of 
a method $\meth \in \Meth$, 
a data value $\fresh \in \domain$, 
an operation $\op \in \Op$, and is denoted 
$\callevent{\meth}{\fresh}{\op}$. 
Similarly, a \emph{return \action{}} is denoted 
$\retevent{\meth}{\fresh}{\op}$. The 
\operation{} $\op$ is used to match return \action{s} to their call \action{s}.

\begin{definition}

  A \emph{(concurrent) \execution} $\exec$ is a sequence of call and return
  \action{s} which satisfy a well-formedness property: every return has a call
  \action{} before it in $\exec$, using the same tuple $\meth,\fresh,\op$, and
  an \operation{} $\op$ can be used only twice in $\exec$, once in a call
  \action{}, and once in a return \action{}.

\end{definition}

\begin{example}

  The sequence $\callevent{\enqueue}{7}{\op_1} \cc
  \callevent{\enqueue}{4}{\op_2}\cc \retevent{\enqueue}{7}{\op_1}\cc
  \retevent{\enqueue}{4}{\op_2}$ is an \execution, while
  $\callevent{\enqueue}{7}{\op_1} \cc \callevent{\enqueue}{4}{\op_2}\cc
  \retevent{\enqueue}{7}{\op_1}\cc \retevent{\enqueue}{4}{\op_1}$ and
  $\callevent{\enqueue}{7}{\op_1} \cc \retevent{\enqueue}{7}{\op_1}\cc
  \retevent{\enqueue}{4}{\op_2}$ are not.

\end{example}

\begin{definition}

  An \emph{\implementation} $\impl$ is a set of (concurrent) \execution{s}.

\end{definition}

Implementations represent libraries whose methods are called by external
programs, giving rise to the following closure
properties~\citep{Bouajjani:2015:TRC:2676726.2677002}. In the following, $c$
denotes a call \action{}, $r$ denotes a return \action{}, $a$ denotes any
\action{}, and $e$, $e'$ denote \execution{s}.
\begin{itemize}

  \item Programs can call library methods at any point in time: \\
  $e \cdot e' \in \impl$ implies $e \cdot c \cdot e' \in \impl$
  so long as $e \cdot c \cdot e'$ is well formed.

  \item Calls can be made earlier: \\
  $e \cdot a \cdot c \cdot e' \in \impl$ implies $e \cdot c \cdot a \cdot e' \in \impl$.

  \item Returns been made later: \\
  $e \cdot r \cdot a \cdot e' \in \impl$ implies $e \cdot a \cdot r \cdot e' \in \impl$.

\end{itemize}
Intuitively, these properties hold because call and return \action{s} are not
visible to the other threads which are running in parallel.

For the remainder of this work, we consider only \emph{completed}
\execution{s}, where each call \action{} has a corresponding return \action{}.
This simplification is sound when implementation methods can always make
progress in isolation~\citep{conf/concur/HenzingerSV13}: formally, for any
execution $e$ with pending operations, there exists an execution $e'$
obtained by extending $e$ only with the return actions of the pending
operations of $e$. Intuitively this means that methods can always return
without any help from outside threads, avoiding deadlock.

We simply reasoning on executions by abstracting them into \emph{histories}.
\begin{definition}

  A \emph{\history} is a \emph{labeled partial order}
  $(\ops,<,\getlabel)$ with $\ops \subseteq \Op$ and $\getlabel: \ops
  \rightarrow \Meth \times \domain$.

\end{definition}
The order $<$ is called the \emph{\happenedbefore{} relation}, and we say that
$\op_1$ \emph{happens before} $\op_2$ when $\op_1 < \op_2$. Since histories
arise from executions, their happens-before relations are \emph{interval
orders}~\citep{Bouajjani:2015:TRC:2676726.2677002}: for distinct $\op_1, \op_2,
\op_3, \op_4$, if $\op_1 < \op_2$ and $\op_3 < \op_4$ then either $\op_1 <
\op_4$, or $\op_3 < \op_2$. Intuitively, this comes from the fact that
concurrent threads share a notion of global time. $\domof{\hist} \subseteq
\domain$ denotes the set of data values appearing in $\hist$.

The \emph{history of an \execution{} $\exec$} is defined
as $(\ops,<,\getlabel)$ where:
\begin{itemize}
\item $\ops$ is the set of \operation{s} which appear in $\exec$,
\item 
  $\op_1 < \op_2$ iff the return \action{} of $\op_1$ is before
  the call \action{} of $\op_2$ in $\exec$,
\item
  an \operation{} $\op$ occurring in a call \action{} 
  $\callevent{\meth}{\fresh}{\op}$ is 
  labeled by $\methevent{\meth}{\fresh}$.
\end{itemize}

\begin{example}
The history of the \execution{} 
$\callevent{\enqueue}{7}{\op_1} \cc
\callevent{\enqueue}{4}{\op_2}\cc
\retevent{\enqueue}{7}{\op_1}\cc
\retevent{\enqueue}{4}{\op_2}$ is 
$(\set{\op_1,\op_2},<,\getlabel)$ with 
$\getlabel(\op_1) = \dataletter{\enqueue}{7}$,
$\getlabel(\op_2) = \dataletter{\enqueue}{4}$, and with
$<$ being the empty order relation, since $\op_1$ and $\op_2$
\emph{overlap}.
\end{example}



Let $\hist = (\ops,<,\getlabel)$ be a history and $\useq$ a \dataword{} of
length $\ulength$. We say that $\hist$ is \emph{linearizable with respect to}
$\useq$, denoted $\linable{\hist}{\useq}$, if there is a bijection \mbox{$\bij:
\ops \rightarrow \set{1,\dots,\ulength}$ s.t.}
\begin{itemize}

  \item if $\op_1 < \op_2$ then $f(\op_1) < f(\op_2)$,

  \item the \methodevent{} at position $f(\op)$ in $\useq$ is $\getlabel(\op)$.

\end{itemize}

\begin{definition}

  A history $\hist$ is \emph{linearizable} with respect to a set $\dwords$ of
  \dataword{s}, denoted $\linable{\hist}{\dwords}$, if there exists $\useq \in
  \dwords$ such that $\linable{\hist}{\useq}$.

\end{definition}
A set of histories $\hists$ is \emph{linearizable} with respect to $\dwords$,
denoted $\linable{\hists}{\dwords}$ if $\linable{\hist}{\dwords}$ for all
$\hist \in \hists$. We extend these definitions to \execution{s} according to
their histories.

A \dataword{} $\useq$ is said to be \emph{\distinguished} if, for all \main{} 
methods $\meth \in \mainmethod$, and every $\fresh \in \domain$, 
there is at most one \methodevent{} $\dataletter{\meth}{\fresh}$ in $\useq$. 
The subset of \distinguished{} \dataword{s} of a set $\dwords$ is denoted by
$\restrict{\dwords}$.
The definition extends to (sets of) \execution{s} and histories. For instance, 
an \execution{} is \distinguished{} if for all \main{} methods
$\meth \in \mainmethod$ and every $\fresh \in \domain$, there is
at most one call \action{} $\callevent{\meth}{\fresh}{\op}$.

\begin{example}
$\callevent{\enqueue}{7}{\op_1} \cc
\callevent{\enqueue}{7}{\op_2}\cc
\retevent{\enqueue}{7}{\op_1} \cc
\retevent{\enqueue}{7}{\op_2}$ is not \distinguished{}, as there are
two call \action{s} with the same \main{} method (\enqueue) and the same
data value.
\end{example}

A \emph{renaming} $r$ is a function from $\domain$ to $\domain$. Given a 
\dataword{} (\resp execution or history) $\useq$, we denote by 
$\rename{\useq}$ the \dataword{} (\resp execution or history) obtained
from $\useq$ by replacing every data value $\fresh$ by $r(\fresh)$. 

\begin{definition}

  The set of \dataword{s} (\resp \execution{s} or histories) $\dwords$ is
  \emph{data independent} if:
  \newcommand{\nuseq}{\useq'}
  \begin{itemize}
  \item 
    for all $\useq \in \dwords$, there exists $\nuseq \in \restrict{\dwords}$, 
    and a renaming $r$ such that $\useq = \rename{\nuseq}$,
  \item for all $\useq \in \dwords$ and for all renaming $r$, 
  $\rename{\useq} \in \dwords$.
  \end{itemize}

\end{definition}

When checking that a data-independent \implementation{} $\impl$ is linearizable
with respect to a data-independent specification $\ds$, it is enough to do
so for \distinguished{} \execution{s}~\citep{conf/tacas/AbdullaHHJR13}.
Thus, in the remainder of the paper, we focus on characterizing 
linearizability for \distinguished{} \execution{s}, rather than
arbitrary ones.

\begin{restatable}[\citet{conf/tacas/AbdullaHHJR13}]{lemma}{dataindependentlemma}
A data-independent implementation $\impl$ is linearizable with respect to a
data-independent specification $\ds$, if and only if $\restrict{\impl}$ is 
linearizable with respect to $\restrict{\ds}$.
\end{restatable}
\section{Inductively-Defined Data Structures}

\label{sec:structures}


A \emph{\datastructure} $\ds$ is given syntactically as an ordered sequence 
of \ruler{s} $\rul_1,\dots,\rul_\nrule$, each of the form
$\useq_1 \cc \useq_2 \cdots \useq_\nseq \in \ds \land 
\conditions(\useq_1,\dots,\useq_\nseq) 
\entails \constructor(\useq_1,\dots,\useq_\nseq) \in \ds$, where
the variables $u_i$ are interpreted over method-event sequences, and
\begin{itemize}
\item 
  $\conditions(\useq_1,\dots,\useq_\nseq)$ is a conjunction of conditions
    on $\useq_1,\dots,\useq_\nseq$ with atoms
  \begin{itemize}
  \item $\useq_i \in \meths^*$ ($\meths \subseteq \Meth$)
  \item $\lesser{0}{\meth}{\useq_i}$
  \end{itemize}
\item 
  $\constructor(\useq_1,\dots,\useq_\nseq)$ is an \emph{expression} 
  $\expr = \fact_1\cc\fact_2\cdots\fact_\nfact$ where
  \begin{itemize}
    \item $\useq_1,\dots,\useq_\nseq$ appear in that order, exactly once,
      in $\expr$,
    \item 
      each $\fact_i$ is either some $\useq_j$, a method $\meth$, or 
      a Kleene closure $\meth^*$ ($\meth \in \Meth$),
    \item 
      a method $\meth \in \Meth$ appears at most once in $\expr$.
  \end{itemize}
\end{itemize}
We allow $\nseq$ to be $0$ for base rules, such as $\emptyseq \in \ds$.

A condition $\useq_i \in \meths^*$ ($\meths \subseteq \Meth$) is satisfied
when the methods used in $\useq_i$ are all in $\meths$. The predicate 
$\lesser{0}{\meth}{\useq_i}$ is satisfied when, for every
\methodevent{} $\methevent{\meth}{\fresh}$ in $\useq_i$, there exists another 
\methodevent{} in $\useq_i$ with the same data value $\fresh$.

Given a \dataword{} 
$\useq = \useq_1 \cc \dots \cc \useq_\nseq$ and an expression 
$\expr = \constructor(\useq_1,\dots,\useq_\nseq)$, we define 
$\dw{\expr}$ as the set of \dataword{s} which can be obtained from
$\expr$ by replacing the methods $\meth$ by a \methodevent{} 
$\methevent{\meth}{\fresh}$ and the Kleene closures $\meth^*$ by $0$ or more 
\methodevent{s} $\methevent{\meth}{\fresh}$. 
All \methodevent{s} must use the same data value $\fresh \in \domain$.

A \ruler{} 
$\rul \equiv\ \useq_1 \cc \useq_2 \cdots \useq_\nseq \in \ds \land 
\conditions(\useq_1,\dots,\useq_\nseq) 
\entails \constructor(\useq_1,\dots,\useq_\nseq) \in \ds$ 
is applied to a \dataword{} $\wseq$ to obtain a 
new \dataword{} $\wseq'$ from the set:
\[
  \bigcup_{\substack{\wseq = \wseq_1 \cc \wseq_2 \cdots \wseq_\nseq \land\\
    \conditions(\wseq_1,\dots,\wseq_\nseq)}}
      \dw{\constructor(\wseq_1,\dots,\wseq_\nseq)}
\]
We denote this $\wseq \applyrule{\rul} \wseq'$. The set of \dataword{s}
$\dw{\ds} = \dw{\rul_1,\dots,\rul_\nrule}$ is then defined as the set of
\dataword{s} $\wseq$ which can be derived from the empty word:
\[
  \emptyseq = \wseq_0 \applyrule{\rul_{i_1}} \wseq_1 \applyrule{\rul_{i_2}}
  \wseq_2 \dots \applyrule{\rul_{i_p}} \wseq_p = \wseq \text{,}
\]
where $i_1,\dots,i_p$ is a non-decreasing sequence of integers from
$\set{1\dots,\nrule}$. This means that the \ruler{s} must be applied in order,
and each \ruler{} can be applied $0$ or several times.

Below we give inductive definitions for the atomic queue and stack
\datastructure{s}. Other data structures such as atomic registers and mutexes
also have inductive definitions, as demonstrated in \iftoggle{fullversion}{the
appendix}{the extended version of this
paper~\citep{DBLP:journals/corr/BouajjaniEEH15}}.

\begin{example}
The queue has a method $\enqueue$ to add an element to the \datastructure{},
and a method $\dequeue$ to remove the elements in a FIFO order. The method 
$\deqempty$ can only return when the queue is empty (its parameter is
not used). The only 
\main{} method is $\enqueue$. 
Formally, $\queue$ is defined by the \ruler{s} 
$\rstart,\renq,\renqdeq$ and $\rempty$.
\footnotesize
\begin{flalign*}
  \rstart \equiv&\ \emptyseq \in \queue\\
  \renq \equiv&\ \useq \in \queue \land \useq \in \oenqueue^* \entails 
    \useq \cc \oenqueue \in \queue\\
\renqdeq \equiv&\ \useq \cc \vseq \in \queue \land 
  \useq \in \enqueue^* \land
  \vseq \in \set{\enqueue,\dequeue}^* 
  \entails
    \enqueue \cc \useq \cc \dequeue \cc \vseq \in \queue\\
\rempty \equiv&\ \useq \cc \vseq \in \queue \land 
  \lesser{0}{\enqueue}{\useq}
  \entails \useq \cc \deqempty \cc \vseq \in 
  \queue
\end{flalign*}
\normalsize
One derivation for $\queue$ is:
\footnotesize
\begin{flalign*}
\emptyseq \in \queue
&\applyrule{\renqdeq} 
  \dataletter{\enqueue}{1} \cc
  \dataletter{\dequeue}{1} \in \queue\\
&\applyrule{\renqdeq} 
  \dataletter{\enqueue}{2} \cc
  \dataletter{\enqueue}{1} \cc
  \dataletter{\dequeue}{2} \cc
  \dataletter{\dequeue}{1}\in \queue \\
&\applyrule{\renqdeq} 
  \dataletter{\enqueue}{3} \cc
  \dataletter{\dequeue}{3} \cc
  \dataletter{\enqueue}{2} \cc
  \dataletter{\enqueue}{1} \cc
  \dataletter{\dequeue}{2} \cc
  \dataletter{\dequeue}{1} \in \queue\\
&\applyrule{\rempty} 
  \dataletter{\enqueue}{3} \cc
  \dataletter{\dequeue}{3} \cc
  \deqempty \cc
  \dataletter{\enqueue}{2} \cc
  \dataletter{\enqueue}{1} \cc
  \dataletter{\dequeue}{2} \cc
  \dataletter{\dequeue}{1} \in \queue
\end{flalign*}
\normalsize
Similarly, 
$\stack$ is composed of the \ruler{s}
$\rstart,\rpushpop,\rpush,\rpopempty$.
\footnotesize
\begin{flalign*}
  \rstart \equiv&\ \emptyseq \in \stack\\
\rpushpop \equiv\ & \useq \cc \vseq \in \stack \land \lesser{0}{\push}{\useq}
    \land \lesser{0}{\push}{\vseq} 
    \land \useq,\vseq \in \set{\push,\pop}^*
    \\ &
    \entails 
    \push \cc \useq \cc \pop \cc \vseq \in \stack\\
\rpush \equiv\ & \useq \cc \vseq \in \stack \land \lesser{0}{\push}{\useq}
    \land \useq,\vseq \in \set{\push,\pop}^*
  \entails   \useq \cc \opush \cc \vseq \in \stack\\
\rpopempty \equiv\ & \useq \cc \vseq \in \stack 
  \land \lesser{0}{\push}{\useq} 
  \entails \useq \cc \popempty \cc \vseq \in \stack
\end{flalign*}
\normalsize
\end{example}

We assume that the rules defining a \datastructure{} $\ds$ satisfy a 
non-ambiguity property stating that the last step in deriving a  \dataword{} 
in $\dw{\ds}$ is unique and it can be effectively determined. Since we are 
interested in characterizing the linearizations of a history and its
projections, this property is extended to permutations of projections of 
\dataword{s} which are admitted by $\ds$. Thus, we assume that the rules 
defining a \datastructure{} are \emph{\wellformed}, that is:

\begin{itemize}
\item 
  for all $\useq\in \dw{\ds}$, there exists a unique \ruler{}, 
  denoted by $\getrule(\useq)$, that can be used as the last
  step to derive $\useq$, i.e., for every sequence of rules 
  $\rul_{i_1},\ldots,\rul_{i_n}$ 
  leading to $\useq$, $\rul_{i_n}=\getrule(\useq)$. 
  For $\useq \not\in \dw{\ds}$, $\getrule(\useq)$ is also defined but can be
  arbitrary, as there is no derivation for $\useq$.
\item 
  if $\getrule(\useq) = \rul_i$, then for every permutation 
  $\useq'\in \dw{\ds}$ of a 
  projection of $u$, $\getrule(\useq') = \rul_j$ with $j \leq i$. If $\useq'$ is a permutation of $u$, 
  then $\getrule(\useq') = \rul_i$.
\end{itemize}


Given a (completed) history $\hist$, all the $\useq$ such that 
$\linable{\hist}{\useq}$ are permutations of one another. 
The last condition of \wellformedness{} thus enables us to extend the function 
$\getrule$ to histories: $\getrule(\hist)$ is defined as $\getrule(\useq)$ 
where $\useq$ is any \dataword{} such that $\linable{\hist}{\useq}$.
We say that $\getrule(\hist)$ is the \ruler{} \emph{\corresponding} to
$\hist$.

\begin{example}
For $\queue$, we define $\getrule$ for a \dataword{} $\useq$ as follows:
\begin{itemize}
\item if $\useq$ contains a $\deqempty$ operation, $\getrule(\useq) = \rempty$,
\item else if $\useq$ contains a $\dequeue$ operation, $\getrule(\useq) = \renqdeq$, 
\item else if $\useq$ contains only $\enqueue$'s, $\getrule(\useq) = \renq$,
\item else (if $\useq$ is empty), $\getrule(\useq) = \rstart$.
\end{itemize}
Since the conditions we use to define $\getrule$ are closed under 
permutations, we get that for any permutation $\useq_2$ of $\useq$,
$\getrule(\useq) = \getrule(\useq_2)$, and $\getrule$ can be extended to
histories. Therefore, the rules $\rstart, \renqdeq, \rempty$ are \wellformed.
\end{example}

\section{Reducing Linearizability to State Reachability}
\label{sec:reduction}

Our end goal for this section is to show that for any data-independent
implementation $\impl$, and any specification $\ds$ satisfying several 
conditions defined in the following, there exists a computable 
finite-state automaton $\aut$ (over call and return \action{s}) such that:
\[
  \linable{\impl}{\ds} \iff \impl \cap \aut = \emptyset
\]
Then, given a model of $\impl$, the linearizability of $\impl$ is reduced
to checking emptiness of the synchronized product between the model of $\impl$
and $\aut$. The automaton $\aut$ represents (a subset of the) \execution{s} 
which are not linearizable with respect to $\ds$.

The first step in proving our result is to show that, under some conditions, we 
can partition the concurrent executions which are not linearizable with 
respect to $\ds$ into a finite number of classes. Intuitively, each 
non-linearizable execution must correspond to a violation for one of the 
rules in the definition of $\ds$.

We identify a property, which we call \emph{step-by-step linearizability},
which is sufficient to obtain this characterization. Intuitively, step-by-step
linearizability enables us to build a linearization for an \execution{} $\exec$
incrementally, using linearizations of projections of $\exec$.

The second step is to show that, for each class of violations (\ie with respect
to a specific \ruler{} $\rul_i$), we can build a regular automaton $\aut_i$ 
such that: a) when restricted to well-formed executions, 
$\aut_i$ recognizes a subset of this class; b) 
each non-linearizable execution has a corresponding execution, 
obtained by data independence, accepted by $\aut_i$.
If such an automaton exists, 
we say that $\rul_i$ is \emph{\tractable} 
(formally defined later in this section).

We prove that, provided these two properties hold, we have the 
equivalence mentioned above, by defining $\aut$ as the union of the 
$\aut_i$'s built for each \ruler{} $\rul_i$.


\subsection{Reduction to a Finite Number of Classes of Violations}

Our goal here is to give a characterization of the \dataword{s} which 
belong to a \datastructure{}, as well as to 
give a characterization of the concurrent \execution{s} which are linearizable with 
respect to the \datastructure{}. This characterization enables us
to classify the linearization violations into a finite number of 
classes.

Our characterization relies heavily 
on the fact that the \datastructure{s} we consider are 
\emph{\closedbyerasure}, 
\ie for all $\useq \in \ds, \dats \subseteq \domain$, we have 
$\project{\useq}{\dats} \in \ds$. The reason for this is that the guards
used in the inductive \ruler{s} are \closedbyerasure{}.

\begin{restatable}{lemma}{elasticity}
\label{lem:elasticity}
Any \datastructure{} $\ds$ defined in our framework is \closedbyerasure{}.
\end{restatable}

A \dataword{} $\useq$ is said to \emph{match} a \ruler{} $\rul$ with conditions 
$\conditions$ if there exist a data value $\fresh$ and 
\dataword{s} $\useq_1,\dots,\useq_\nseq$ such that 
$\useq$ can be written as $\dw{\constructor(\useq_1,\dots,\useq_\nseq)}$, 
where $\fresh$ is the data value used for the \methodevent{s}, and such
that $\conditions(\useq_1,\dots,\useq_\nseq)$ holds. We call $\fresh$ the
\emph{witness} of the decomposition.
We denote by $\matchset{\rul}$ the set of \dataword{s} which match $\rul$,
and we call it the \emph{\matchingset} of $\rul$.

\begin{example}
$\matchset{\renqdeq}$ is the set of \dataword{s} of the form
$\dataletter{\enqueue}{\fresh} \cc \useq \cc 
\dataletter{\dequeue}{\fresh} \cc \vseq$ for some $\fresh \in \domain$,
and with $\useq \in \enqueue^*$.
\end{example}

\begin{restatable}{lemma}{characterizationseq}
\label{lem:characterizationseq}
Let $\ds = \rul_1,\dots,\rul_\nrule$ be a \datastructure{} and $\useq$  a \distinguished{} \dataword{}. Then,
\[
  \useq \in \ds \iff 
    \projections{\useq} \subseteq \bigcup_{i \in \set{1,\dots,\nrule}}
    \matchset{\rul_i}
\]
\end{restatable}

This characterization enables us to get rid of the recursion, so that we only 
have to check non-recursive properties. We want a similar lemma to
characterize $\linable{\exec}{\ds}$ for an \execution{} $\exec$. This is where 
we introduce the notion of \emph{step-by-step linearizability}, as the lemma
will hold under this condition.

\begin{definition}
A \datastructure{} $\ds = \rul_1,\dots,\rul_\nrule$ is 
said be to \emph{\sbs} if for any \distinguished{} \execution{} $\exec$, 
if $\exec$ is linearizable w.r.t. $\matchset{\rul_i}$ with 
witness $\fresh$, we have:
\[
\linable{\removefrom{\exec}{\fresh}}
{\dw{\rul_1,\dots,\rul_i}} \implies
\linable{\exec}{\dw{\rul_1,\dots,\rul_i}}
\]

\end{definition}


This notion applies to the usual \datastructure{s}, as shown by the following
lemma.
The generic schema we use is the following:
we let $\useq' \in \dw{\rul_1,\dots,\rul_i}$ be a \dataword{} such that 
$\linable{\removefrom{\exec}{\fresh}}{\useq'}$ and build a
graph $G$ from $\useq'$, whose acyclicity implies that 
$\linable{\exec}{\dw{\rul_1,\dots,\rul_i}}$. Then, we show that we can always
choose $\useq'$ so that $G$ is acyclic. 

\begin{restatable}{lemma}{monoobs}
\label{lem:monoobs}
$\queue$,
$\stack$, 
$\register$,
and $\mutex$ 
are \sbs.
\end{restatable}

Intuitively, step-by-step linearizability will help us prove the 
right-to-left direction of \lem{lem:characterizationpo} by allowing us
to build a linearization for $\exec$ incrementally, from the linearizations
of projections of $\exec$.

\begin{restatable}{lemma}{characterizationpo}
\label{lem:characterizationpo}
Let $\ds$ be a \datastructure{} with \ruler{s} 
$\rul_1,\dots,\rul_\nrule$.
Let $\exec$ be a \distinguished{} \execution{}. 
If $\ds$ is \sbs, we have (for any $j$):
\[
  \linable{\exec}{\dw{\rul_1,\dots,\rul_j}} \iff 
    \linable{\projections{\exec}}{\bigcup_{i \leq j} \matchset{\rul_i}}
\]
\end{restatable}

Thanks to \lem{lem:characterizationpo}, if we're looking for an \execution{} 
$\exec$ which is not linearizable w.r.t. some data-structure $\ds$, we 
must prove that 
$\notlinable{\projections{\exec}}{\bigcup_i \matchset{\rul_i}}$, \ie 
we must find a projection $\exec' \in \projections{\exec}$ which is not 
linearizable with respect to any $\matchset{\rul_i}$
($\notlinable{\exec'}{\bigcup_i \matchset{\rul_i}}$).

This is challenging as it is difficult to check that an \execution{} is not 
linearizable w.r.t. a union of sets simultaneously. 
Using \wellformedness, we simplify this check by making it more modular, so 
that we only have to check one set $\matchset{\rul_i}$ at a time.

\begin{restatable}{lemma}{exclu}
\label{lem:exclu}
Let $\ds$ be a \datastructure{} with \ruler{s} 
$\rul_1,\dots,\rul_\nrule$.
Let $\exec$ be a \distinguished{} \execution{}. 
If $\ds$ is \sbs, we have:
\[
  \linable{\exec}{\ds} \iff 
    \forall \exec' \in \projections{\exec}.\ 
    \linable{\exec'}{\matchset{\rul}}
	\text{ where }\rul = \getrule(\exec')
\]
\end{restatable}

\lem{lem:exclu} gives us the finite kind of violations that we mentioned in
the beginning of the section. More precisely, if we negate both sides of 
the equivalence, we have:
$
  \notlinable{\exec}{\ds} \iff 
    \exists \exec' \in \projections{\exec}.\ 
    \notlinable{\exec'}{\matchset{\rul}}
$.
This means that whenever an \execution{} is not linearizable w.r.t. 
$\ds$, there can be only finitely reasons, namely there must exist a projection
which is not linearizable w.r.t. the matching set of its 
corresponding rule.

\subsection{Regularity of Each Class of Violations}

Our goal is now to construct, for each $\rul$, an automaton $\aut$ which 
recognizes (a subset of) the executions $\exec$, which have a projection 
$\exec'$ such that $\notlinable{\exec'}{\matchset{\rul}}$.
More precisely, we want the following property.

\begin{definition}
A \ruler{} $\rul$ is said to be \emph{\tractable} if we can build an automaton
$\aut$ such that, for any \dataindependent{} \implementation{} $\impl$, we 
have:
\[
  \aut \cap \impl \neq \emptyset \iff
  \exists \exec \in \restrict{\impl}, \exec' \in \projections{\exec}.\ 
    \getrule(\exec') = \rul \land \notlinable{\exec'}{\matchset{\rul}}
\]
A \datastructure{} $\ds$ is \emph{\tractable} if all of its rules are
\coregular.
\end{definition}

Formally, the alphabet of $\aut$ is
$
\set{\callevent{\meth}{\fresh}{}\ |\ \meth \in \Meth, \fresh \in \dats} \cup
\set{\retevent{\meth}{\fresh}{}\ |\  \meth \in \Meth, \fresh \in \dats}
$
for a finite subset $\dats \subseteq \domain$. The automaton doesn't read
operation identifiers, thus, when taking the intersection with $\impl$, we
ignore them.

\begin{figure}[t]
  \begin{minipage}[t]{0.40\textwidth}
    \centering
    \begin{tikzpicture}[scale=0.5]
\scriptsize

\newcommand{\makeinterval}[4]{
  \node at (#2,#1) (a) {};
  \node at (#3,#1) (b) {};
%
%
  \draw[|-|] (a) -- coordinate (d) (b);
  
  \node[above] at (d) {#4};
}

\makeinterval{0}{2.5}{6.5}{$\dataletter{\deqempty}{2}$}
\makeinterval{-1}{0}{2}{$\dataletter{\enqueue}{1}$}
\makeinterval{-2}{1}{3}{$\dataletter{\enqueue}{1}$}
\makeinterval{-3}{0.5}{4.2}{$\dataletter{\enqueue}{1}$}
\makeinterval{-4}{2}{5.5}{$\dataletter{\enqueue}{1}$}

\makeinterval{-1}{3.5}{8}{$\dataletter{\dequeue}{1}$}
\makeinterval{-2}{4.4}{7.4}{$\dataletter{\dequeue}{1}$}
\makeinterval{-3}{5.7}{7.8}{$\dataletter{\dequeue}{1}$}
\makeinterval{-4}{7}{8.2}{$\dataletter{\dequeue}{1}$}

\end{tikzpicture}
    \caption{A four-pair $\rempty$ violation.
    \iftoggle{fullversion}{\lem{lem:rempty-tract}}{The extended version of this paper}
    demonstrates that this pattern with arbitrarily-many pairs is regular.}
    \label{fig:intervals}
  \end{minipage}
  \hfill
  \begin{minipage}[t]{0.56\textwidth}
    \centering
    \begin{tikzpicture}[x=2.0cm, y=0.4cm, ->]
\scriptsize
\node[state, initial, initial text=] (a) {$q_0$};
\node[state,above=1 of a] (b) { $q_1$ };
\node[state,right=1 of b] (c) { $q_2$ };
\node[state,below=1 of c] (d) { $q_3$ };
\node[state, accepting, right=1 of c] (e) { $q_4$ };

\path (a) edge [loop right] node[right] {$\thealphabet{3}$} (a);
\path (b) edge [loop above] node[above] {$\thealphabet{3}$} (b);
\path (c) edge [loop above] node[above] {$\thealphabet{3}$} (c);
\path (d) edge [loop right] node[right] {$\thealphabet{3}$} (d);
\path (e) edge [loop above] node[above] {$\thealphabet{3}$} (e);

\path (a) edge [loop below] node[below] {$\callevent{\enqueue}{1}{}$} (a);
\path (a) edge node[right,yshift=-2] { $\retevent{\enqueue}{1}{}$ } (b);
\path (b) edge node[above] { $\callevent{\deqempty}{2}{}$ } (c);
\path (c) edge node[above] { $\retevent{\deqempty}{2}{}$ } (e);
\path (c) edge [bend left] node[right] { $\retevent{\enqueue}{1}{}$ } (d);
\path (d) edge [bend left] node[left,yshift=6] { $\callevent{\dequeue}{1}{}$ }  (c);

\end{tikzpicture}
    \caption{An automaton recognizing $\rempty$ violations, for which the queue
    is non-empty, with data value $1$, for the span of $\deqempty$. We assume
    all $\callevent{\enqueue}{1}{}$ actions occur initially without loss of
    generality due to implementations’ closure properties.}
    \label{fig:queue-empty}
  \end{minipage}
\end{figure}

\begin{restatable}{lemma}{regularity}
\label{lem:fourregular}
$\queue$, $\stack$, $\register$, and $\mutex$
are \coregular.
\end{restatable}
\begin{proof}
To illustrate this lemma, we sketch the proof for the \ruler{} $\rempty$ of 
$\queue$. The complete proof of the lemma can be found in the extended
version of this paper.

We prove in \iftoggle{fullversion}{the appendix (\coro{coro:cover})}{the
extended version} that a history has a projection such that
$\getrule(\hist') = \rempty$ and $\notlinable{\hist'}{\matchset{\rempty}}$ if
and only if it has a $\deqempty$ operation which is \emph{covered} by other
operations, as depicted in \fig{fig:intervals}. The automaton $\autof{\rempty}$
in \fig{fig:queue-empty} recognizes such violations.

Let $\impl$ be any data-independent implementation. We show that
\[
  \autof{\rempty} \cap \impl \neq \emptyset \iff
  \exists \exec \in \restrict{\impl}, \exec' \in \projections{\exec}.\ 
    \getrule(\exec') = \rempty \land \notlinable{\exec'}{\matchset{\rempty}}
\]
$(\Rightarrow)$ Let $\exec \in \impl$ be an execution which is accepted
by $\autof{\rempty}$.
By data independence, let $\restrict{\exec} \in \impl$ 
and $r$ a renaming such that $\exec = r(\restrict{\exec})$. Let 
$\dat_1,\dots,\dat_\ndats$ be the data values which are mapped to value $1$ by 
$r$. 

Let $\dat$ be the data value which is mapped to value $2$ by $r$. Let
$\theempty$ the $\deqempty$ operation with data value $\dat$. By construction
of the automaton we can prove that $\theempty$ is covered by
$\dat_1,\dots,\dat_\ndats$, and \iftoggle{fullversion}{using
\coro{coro:cover}, }{}conclude that $\hist$ has a projection such that
$\getrule(\hist') = \rempty$ and $\notlinable{\hist'}{\matchset{\rempty}}$.

$(\Leftarrow)$ Let $\restrict{\exec} \in \restrict{\impl}$ such that there is a
projection $\exec'$ such that $\getrule(\exec') = \rempty$ and
$\notlinable{\exec'}{\matchset{\rempty}}$. Let $\dat_1,\dots,\dat_\ndats$ be
the data values given by \iftoggle{fullversion}{\coro{coro:cover}}{the
$\rempty$-characterization in the full version of this paper}, and let $\dat$
be the data value corresponding to the $\deqempty$ operation.

Without loss of generality, we can always choose the cycle so that 
$\dataletter{\enqueue}{\dat_i}$ doesn't happen before 
$\dataletter{\dequeue}{\dat_{i-2}}$ (if it does, drop $\dat_{i-1}$).

Let $r$ be the renaming which maps $\dat_1,\dots,\dat_\ndats$ to $1$, 
$\dat$ to $2$, and all other values to $3$. Let $\exec = r(\restrict{\exec})$. 
The execution $\exec$ can be recognized by automaton $\autof{\rempty}$, and 
belongs to $\impl$ by data independence.
\end{proof}


When we have a \datastructure{} which is both \sbs{} and \coregular{}, we can 
make a linear time reduction from the verification of linearizability with 
respect to $\ds$ to a reachability problem, as illustrated in
\thm{thm:reachability}.

\begin{restatable}{theorem}{reachability}
\label{thm:reachability}
Let $\ds$ be a \sbs{} and \coregular{} \datastructure{} and
let $\impl$ be a \dataindependent{} \implementation. There exists a
regular automaton $\aut$ such that:
\[
  \linable{\impl}{\ds} \iff \impl \cap \aut = \emptyset
\]
\end{restatable}


\section{Decidability and Complexity of Linearizability}
\label{sec:decidability}

Theorem~\ref{thm:reachability} implies that the linearizability problem with
respect to any step-by-step linearizable and \coregular{} specification is
decidable for any data-independent implementation for which checking the
emptiness of the intersection with finite-state automata is decidable. Here,
we give a class $\ourclass$ of data-independent implementations for which the
latter problem, and thus linearizability, is decidable.

Each method of an implementation in $\ourclass$ manipulates a finite number of
local variables which store Boolean values, or data values from $\domain$.
Methods communicate through a finite number of shared variables that also store
Boolean values, or data values from $\domain$. Data values may be assigned, but
never used in program predicates (e.g., in the conditions of \texttt{if} and
\texttt{while} statements) so as to ensure data independence. This class
captures typical implementations, or finite-state abstractions thereof, e.g.,
obtained via predicate abstraction.

Let $\impl$ be an implementation from class $\ourclass$. The automata $\aut$
constructed in the proof of \lem{lem:fourregular} use only data values $1$,
$2$, and $3$. Checking emptiness of $\impl \cap \aut$ is thus equivalent to
checking emptiness of $\notall{\impl}{3} \cap \aut$ with the three-valued
implementation $\notall{\impl}{3} = \set{ \exec \in \impl \mid \exec =
\project{\exec}{\set{1,2,3}} }$. The set $\notall{\impl}{3}$ can be represented
by a Petri net since bounding data values allows us to represent each thread
with a finite-state machine. Intuitively, each token in the Petri net
represents another thread. The number of threads can be unbounded since the
number of tokens can. Places count the number of threads in each control
location, which includes a local-variable valuation. Each shared variable also
has one place per value to store its current valuation.

Emptiness of the intersection with regular automata reduces to the
EXPSPACE-complete coverability problem for Petri nets. Limiting verification to
a bounded number of threads lowers the complexity of coverability to
PSPACE~\citep{petrinets}. The hardness part of \thm{th:complexity} comes
from the hardness of state reachability in finite-state concurrent programs.

\begin{theorem}
\label{th:complexity}
  Verifying linearizability of an implementation in $\ourclass$ with respect
  to a step-by-step linearizable and co-regular specification is 
  PSPACE-complete for a fixed number of threads, and EXPSPACE-complete
  otherwise.
\end{theorem}




\section{Related Work}
\label{sec:related}

Several works investigate the theoretical limits of linearizability
verification. Verifying a single execution against an arbitrary ADT specification is
NP-complete~\cite{journals/siamcomp/GibbonsK97}. Verifying all executions of a
finite-state implementation against an arbitrary ADT specification (given as a
regular language) is EXPSPACE-complete when program threads are
bounded~\cite{journals/iandc/AlurMP00,DBLP:journals/corr/Hamza14}, and
undecidable otherwise~\cite{conf/esop/BouajjaniEEH13}.

Existing automated methods for proving linearizability of an atomic object
implementation are also based on reductions to safety
verification~\cite{conf/tacas/AbdullaHHJR13, conf/concur/HenzingerSV13,
conf/cav/Vafeiadis10}. Vafeiadis~\cite{conf/cav/Vafeiadis10} considers
implementations where operation's \emph{linearization points} are fixed to
particular source-code locations. Essentially, this approach instruments the
implementation with ghost variables simulating the ADT specification at
linearization points. This approach is incomplete since not all implementations
have fixed linearization points. Aspect-oriented
proofs~\cite{conf/concur/HenzingerSV13} reduce linearizability to the
verification of four simpler safety properties. However, this approach has only
been applied to queues, and has not produced a fully automated
and complete proof technique. Dodds et
al.~\cite{Dodds:2015:SCT:2676726.2676963} prove linearizability of stack
implementations with an automated proof assistant. Their approach does not lead
to full automation however, e.g.,~by reduction to safety verification.

\section{Conclusion}

We have demonstrated a linear-time reduction from linearizability for fixed ADT
specifications to control-state reachability, and the application of this
reduction to atomic queues, stacks, registers, and mutexes. Besides yielding
novel decidability results, our reduction enables the use of existing
safety-verification tools for linearizability. While this work only applies the
reduction to these four objects, our methodology also applies to other typical
atomic objects including semaphores and sets. Although this methodology
currently does not capture priority queues, which are not data independent, we
believe our approach can be extended to include them. We leave this for future
work.

  \bibliographystyle{abbrvnat}
  \bibliography{violin}

  \iftoggle{fullversion}{
    \newpage
    \section{Appendix}

\subsection{Examples}
\label{app:structures}

For all examples, the domain $\domain$ is the set of natural numbers $\mathbb{N}$. 

\subsubsection{Stack} 
%
%
Definition of the function $\getrule$ for a \dataword{} $\useq$:
\begin{itemize}
\item 
  if $\useq$ contains a $\popempty$ operation, $\getrule(\useq) = \rpopempty$,
\item 
  else if $\useq$ contains an unmatched $\push$ operation, 
  $\getrule(\useq) = \rpush$, 
\item 
  else if $\useq$ contains a $\pop$ operation,
  $\getrule(\useq) = \rpushpop$,
\item else (if $\useq$ is empty), $\getrule(\useq) = \rstart$.
\end{itemize}

\subsubsection{Register}

The register has a method $\writevar$ used to write a data-value, and 
a method $\readvar$ which returns the last written value. The only 
\main{} method is $\writevar$. Its rules are
$\rstart$ and $\rreg$:
\begin{flalign*}
  \rstart \equiv&\ \emptyseq \in \register\\
\rreg \equiv&\ \useq \in \register \entails
    \writevar \cc \readvar^* \cc \useq \in \register
\end{flalign*}
Definition of the function $\getrule$ for a \dataword{} $\useq$:
\begin{itemize}
\item 
  if $\useq$ is not empty, $\getrule(\useq) = \rreg$,
\item else, $\getrule(\useq) = \rstart$.
\end{itemize}

\subsubsection{Mutex (Lock)}

The mutex has a method $\lock$, used to take ownership of the $\mutex$,
and a method $\unlock$, to release it. The only \main{} method 
is $\lock$. It is composed of the rules
$\rstart,\rlock$ and $\rlockunlock$:
\begin{flalign*}
  \rstart \equiv&\ \emptyseq \in \mutex\\
\rlock \equiv\ &\olock \in \mutex \\
\rlockunlock \equiv\ & 
  \useq \in \mutex \entails \lock \cc \unlock \cc \useq 
    \in \mutex
\end{flalign*}

In practice, $\lock$ and $\unlock$ methods do not have a parameter. Here, the
parameter represents a \emph{ghost variable} which helps us relate 
$\unlock$ to their corresponding $\lock$. Any implementation will be 
data independent with respect to these ghost variables.

Definition of the function $\getrule$ for a \dataword{} $\useq$:
\begin{itemize}
\item 
  if $\useq$ contains an $\unlock$ operation, $\getrule(\useq) = \rlockunlock$,
\item
  else if $\useq$ is not empty, $\getrule(\useq) = \rlock$,
\item 
  else, $\getrule(\useq) = \rstart$.
\end{itemize}

\subsection{Proofs of \sect{sec:reduction}}

\dataindependentlemma*

\begin{proof}
$(\Rightarrow)$ Let $\exec$ be a (\distinguished) execution in $\restrict{\impl}$.
By assumption, it is linearizable with respect to a \dataword{} $\useq$ in $\ds$, and the 
bijection between the operations of $\exec$ and the \methodevent{s} of $\useq$,
ensures that $\useq$ is \distinguished{} and belongs to  $\restrict{\ds}$.

$(\Leftarrow)$ Let $\exec$ be an execution in $\impl$. By data independence of 
$\impl$, we know there exists $\restrict{\exec} \in \restrict{\impl}$ and a 
renaming $r$ such that $r(\restrict{\exec}) = \exec$.
By assumption, $\restrict{\exec}$ is linearizable with respect to a \dataword{} 
$\restrict{\useq} \in \restrict{\ds}$.
We define $\useq = r(\restrict{\useq})$, and know by data independence of $\ds$ 
that $\useq \in \ds$.
Moreover, we can use the same bijection used for 
$\linable{\restrict{\exec}}{\restrict{\useq}}$ to prove that 
$\linable{\exec}{\useq}$.
\end{proof}

\elasticity*

\begin{proof}
Let $\useq \in \ds$ and let $\dats \subseteq \domain$. Since $\useq \in \ds$,
there is a sequence of applications of \ruler{s} starting from the empty
word $\emptyseq$ which can derive $\useq$. We remove from this derivation
all the \ruler{s} corresponding to a data-value $\fresh \notin \dats$, and we 
project all the \dataword{s} appearing in the derivation on the $\dats$.
Since the predicates which appear in the conditions are all \elastic{}, the 
derivation remains valid, and proves that $\project{\useq}{\dats} \in \ds$.
\end{proof}

\characterizationseq*

\begin{proof}
($\Rightarrow$) 
Using \lem{lem:elasticity}, we know that $\ds$ is 
\closedbyerasure{}. Thus, any projection of a \dataword{} $\useq$ of $\ds$ is
itself in $\ds$ and has to match one of the \ruler{s} $\rul_1,\dots,\rul_\nrule$.

($\Leftarrow$) By induction on the size of $\useq$. We know 
$\useq \in \projections{\useq}$, so it can be decomposed to satisfy the 
conditions $\conditions$ of some rule $\rul$ of $\ds$. The recursive 
condition is then verified by induction.
\end{proof}

\characterizationpo*

\begin{proof}
($\Rightarrow$) 
We know there exists $\useq \in \ds$ such that 
$\linable{\exec}{\useq}$. Each projection $\exec'$ of $\exec$ can be 
linearized with respect to some projection $\useq'$ of $\useq$, which
belongs to $\bigcup_i \matchset{\rul_i}$ according to 
\lem{lem:characterizationseq}.

($\Leftarrow$) By induction on the size of $\exec$. We know 
$\exec \in \projections{\exec}$ so it can be linearized with respect to a 
\dataword{} $\useq$ matching some \ruler{} $\rul_k$ ($k < j$) with some witness 
$\fresh$. Let $\exec' = \removefrom{\exec}{\fresh}$.

Since $\ds$ is \wellformed, we know that no projection of $\exec$ can be
linearized to a \matchingset{} $\matchset{\rul_i}$ with $i > k$, 
and in particular no projection of $\exec'$. Thus, we deduce that
$\linable{\projections{\exec'}}{\bigcup_{i \leq k} \matchset{\rul_i}}$, and
conclude by induction that $\linable{\exec'}{\dw{\rul_1,\dots,\rul_k}}$.

We finally use the fact that $\ds$ is \sbs{} to deduce that 
$\linable{\exec}{\dw{\rul_1,\dots,\rul_k}}$ and 
$\linable{\exec}{\dw{\rul_1,\dots,\rul_j}}$ because $k < j$.
\end{proof}

\exclu*

\begin{proof}
$(\Rightarrow)$ Let $\exec' \in \projections{\exec}$. 
By \lem{lem:characterizationpo}, we know that $\exec'$ is linearizable with 
respect to $\matchset{\rul_i}$ for some $i$.
Since $\ds$ is \wellformed, $\getrule(\exec')$ is the only \ruler{} such that 
$\linable{\exec'}{\matchset{\rul}}$ can hold, which ends this part of the 
proof.

$(\Leftarrow)$ Particular case of \lem{lem:characterizationpo}.
\end{proof}

\reachability*

\begin{proof}
Let $\aut_1,\dots,\aut_\nrule$ be the regular automata used to show that 
$\rul_1,\dots,\rul_\nrule$ are \tractable, and let $\aut$ be the 
(non-deterministic) union of the $\aut_i$'s.

$(\Rightarrow)$
Assume there exists an execution $\exec \in \impl \cap \aut$. For some $i$,
$\exec \in \aut_i$.
From the definition of ``\tractable'', we deduce that there exists 
$\exec' \in \projections{\exec}$ such that 
$\notlinable{\exec'}{\matchset{\rul_i}}$, where $\rul_i$ is the \ruler{}
\corresponding{} to $\exec'$. By \lem{lem:exclu}, $\exec$ is not linearizable
with respect to $\ds$.

$(\Leftarrow)$ Assume there exists an execution $\exec \in \impl$ which is 
not linearizable with respect to $\ds$. By \lem{lem:exclu}, it has a 
projection $\exec' \in \projections{\exec}$ such that 
$\notlinable{\exec'}{\matchset{\rul_i}}$, where $\rul_i$ is the \ruler{} 
\corresponding{} to $\exec'$. By definition of ``\tractable'', this means
that $\impl \cap \aut_i \neq \emptyset$, and that 
$\impl \cap \aut \neq \emptyset$.
\end{proof}

\subsection{Step-by-step Linearizability}

\monoobs*

\begin{proof}
Even though we do not have a unique proof that the \datastructure{s} are 
\sbs{}, we have a model of proof which is generic, which we use for each
\datastructure{}.  
The generic schema we use is the following:
we let $\useq' \in \dw{\rul_1,\dots,\rul_i}$ be a \dataword{} such that 
$\linable{\removefrom{\hist}{\fresh}}{\useq'}$ and build a
graph $G$ from $\useq'$, whose acyclicity implies that 
$\linable{\hist}{\dw{\rul_1,\dots,\rul_i}}$. Then we show that we can always
choose $\useq'$ so that this $G$ is acyclic.

For better readability we make a sublemma per \datastructure{}.

\begin{lemma}
\label{lem:queuemonosbs}
\queue{} is \sbs.
\end{lemma}
\begin{proof}
Let $\hist$ be a \distinguished{} history, and $\useq$ a \dataword{} 
such that $\linable{\hist}{\useq}$. We have three cases to consider:

1) $\useq$ matches $\renq$ with witness $\fresh$:
let $\hist' = \removefrom{\hist}{\fresh}$ and assume 
$\linable{\hist'}{\dw{\rstart,\renq}}$. Since $\useq$ matches $\renq$, we know 
$\hist$ only contain $\oenqueue$ operations. 
The set $\dw{\rstart,\renq}$ is composed of the \dataword{s} formed by repeating the 
$\oenqueue$ \methodevent{s}, which means that  
$\linable{\hist}{\dw{\rstart,\renq}}$.

2) $\useq$ matches $\renqdeq$ with witness $\fresh$:
let $\hist' = \removefrom{\hist}{\fresh}$ and assume 
$\linable{\hist'}{\dw{\rstart,\renq,\renqdeq}}$. 
Let $\useq' \in \dw{\rstart,\renq,\renqdeq}$
such that $\linable{\hist'}{\useq'}$. We define a graph $G$ whose nodes are 
the operations of $\hist$ and there is an edge from operation $\op_1$ 
to $\op_2$ if
\begin{enumerate}
\item $\op_1$ happens-before $\op_2$ in $\hist$, \label{edgehb}
\item 
  the \methodevent{} corresponding to $\op_1$ in $\useq'$ is before the one 
  corresponding to $\op_2$, \label{edgeind}
\item
  $\op_1 = \dataletter{\enqueue}{\fresh}$ and $\op_2$ is any other operation,
    \label{edgeenq}
\item
  $\op_1 = \dataletter{\dequeue}{\fresh}$ and $\op_2$ is any other $\dequeue$
  operation. \label{edgedeq} 
\end{enumerate}
If $G$ is acyclic, any total order compatible with $G$ forms a sequence 
$\useq_2$ such that $\linable{\hist}{\useq_2}$ and such that $\useq_2$ can
be built from $\useq'$ by adding $\dataletter{\enqueue}{\fresh}$ at the 
beginning and $\dataletter{\dequeue}{\fresh}$ before all $\dequeue$ 
\methodevent{s}.
Thus, $\useq_2 \in \dw{\rstart,\renq,\renqdeq}$ and 
$\linable{\hist}{\dw{\rstart,\renq,\renqdeq}}$.

Assume that $G$ has a cycle, and consider a cycle $C$ of minimal size. We show 
that there is only one kind of cycle possible, and that this cycle can be 
avoided by choosing $\useq'$ appropriately.
Such a cycle can only contain one \happenedbefore{} edge (edges of 
type~\ref{edgehb}), because if there were 
two, we could apply the interval order property to reduce the cycle. Similarly,
since the order imposed by $\useq'$ is a total order, it also satisfies the
interval order property, meaning that $C$ can only contain one edge of 
type~\ref{edgeind}.

Moreover, $C$ can also contain only one edge of type~\ref{edgeenq}, otherwise 
it would have to go through $\dataletter{\enqueue}{\fresh}$ more than once. 
Similarly, it can contain only one edge of type~\ref{edgedeq}.
It cannot contain a type~\ref{edgeenq} edge
$\dataletter{\enqueue}{\fresh} \rightarrow \op_1$ at the same time as 
a type~\ref{edgedeq} edge $\dataletter{\dequeue}{\fresh} \rightarrow \op_2$,
because we could shortcut the cycle by a type~\ref{edgeenq} edge 
$\dataletter{\enqueue}{\fresh} \rightarrow \op_2$.

Finally, it cannot be a cycle of size $2$. For instance, a type~\ref{edgeind} 
edge cannot form a cycle with a type~\ref{edgehb} edge because
$\linable{\hist'}{\useq'}$.
The only form of cycles left are the two cycles of size $3$ where:
\begin{itemize}
\item $\dataletter{\enqueue}{\fresh}$ is before $\op_1$ 
  (type~\ref{edgeenq}),  $\op_1$ is before $\op_2$ in $\useq'$
  (type~\ref{edgeind}), and $\op_2$ happens-before 
  $\dataletter{\enqueue}{\fresh}$: this is not possible, because 
  $\hist$ is linearizable with respect to $\useq$ which matches $\renqdeq$
  with $\fresh$ as a witness. This means that $\useq$ starts with the 
  \methodevent{}
  $\dataletter{\enqueue}{\fresh}$, and that no operation can happen-before 
  $\dataletter{\enqueue}{\fresh}$ in $\hist$.
\item 
  $\dataletter{\dequeue}{\fresh}$ is before $\op_1$ 
  (type~\ref{edgedeq}), $\op_1$ is before $\op_2$ in $\useq'$
  (type~\ref{edgeind}), and $\op_2$ happens-before 
  $\dataletter{\dequeue}{\fresh}$: by definition, we know that 
  $\op_1$  is a $\dequeue$ operation; moreover, since $\hist$ is linearizable 
  with respect to $\useq$ which matches $\renqdeq$ with $\fresh$ as a witness,
  no $\dequeue$ operation can happen-before $\dataletter{\dequeue}{\fresh}$
  in $\hist$, and $\op_2$ is an $\enqueue$ operation (or $\oenqueue$).
  Let $\dat_1,\dat_2 \in \domain$ such that 
  $\dataletter{\dequeue}{\dat_1} = \op_1$ and 
  $\dataletter{\enqueue}{\dat_2} = \op_2$. 
  
  Since $\op_1$ is before $\op_2$ in $\useq'$,
  we know that $\dat_1$ and $\dat_2$ must be different. Moreover, there is no
  \happenedbefore{} edge from $\op_1$ to $\op_2$, or otherwise, by transitivity
  of the \happenedbefore{} relation, we'd have a cycle of size $2$ between 
  $\op_1$ and $\dataletter{\dequeue}{\fresh}$.
  
  Assume without loss of generality that $\op_1$ is the rightmost $\dequeue$
  \methodevent{} which is before $\op_2$ in $\useq'$, and let 
  $\op_2^1,\dots,\op_2^\nops$ be the $\enqueue$ (or $\oenqueue$) 
  \methodevent{s}
  between $\op_1$ and $\op_2$. There is no \happenedbefore{} edge 
  $\op_1 \hb \op_2^i$, because by applying the interval order property 
  with the other \happenedbefore{} edge 
  $\op_2 \hb \dataletter{\dequeue}{\fresh}$, we'd either have
  $\op_1 \hb \dataletter{\dequeue}{\fresh}$ (forming a cycle of size $2$)
  or $\op_2 \hb \op_2^i$ (not possible because $\linable{\hist'}{\useq'}$ and
  $\op_2^i$ is before $\op_2$ in $\useq'$).
  
  Let $\useq'_2$ be the sequence $\useq'$ where $\dataletter{\dequeue}{\fresh}$
  has been moved after $\op_2$. Since we know there is no \happenedbefore{}
  edge from $\dataletter{\dequeue}{\fresh}$ to $\op_2^i$ or to $\op_2$,
  we can deduce that: $\linable{\hist'}{\useq'_2}$. Moreover, if we consider
  the sequence of deductions which proves that 
  $\useq' \in \dw{\rstart,\renq,\renqdeq}$, we can alter it when we insert the
  pair $\dataletter{\enqueue}{\dat_1}$ and 
  $\op_1 = \dataletter{\dequeue}{\dat_1}$ by inserting $\op_1$ after 
  the $\op_2^i$'s and after $\op_2$, instead of before (the conditions of the
  rule $\renqdeq$ allow it).
\end{itemize}
This concludes case 2), as we're able to choose $\useq'$ so that $G$ is 
acyclic, and prove that $\linable{\hist}{\dw{\rstart,\renq,\renqdeq}}$.

3) $\useq$ matches $\rempty$ with witness $\fresh$:
let $\theempty$ be the $\deqempty$ operation corresponding to the witness.
Let $\hist' = \removefrom{\hist}{\fresh}$ and assume 
$\linable{\hist'}{\queue}$.  Let $L$ be the set of
operations which are before $\theempty$ in $\useq$, and $R$ the ones which are
after. Let $\dats_L$ be the data-values appearing in $L$ and $\dats_R$ be the
data-values appearing in $R$. Since $\useq$ matches $\rempty$, we know that 
$L$ contains no unmatched \enqueue{} operations.

Let $\useq' \in \queue$ such that 
$\linable{\hist'}{\useq'}$.
Let $\useq'_L = \project{\useq'}{\dats_L}$ and 
$\useq'_R = \project{\useq'}{\dats_R}$. 
Since $\queue$ is \closedbyerasure{}, 
$\useq'_L,\useq'_R \in \queue$. Let  
$\useq_2 = \useq'_L \cc \theempty \cc \useq'_R$. 
We can show that $\useq_2 \in \queue$ by using
the derivations of $\useq'_L$ and $\useq'_R$. Intuitively, this is because
$\queue$ is closed under concatenation when the left-hand \dataword{} 
has no unmatched \enqueue{} \methodevent{}, like $\useq'_L$. 

Moreover, we have $\linable{\hist}{\useq_2}$, as shown in the following.
We define a graph $G$ whose nodes are 
the operations of $\hist$ and there is an edge from operation $\op_1$ 
to $\op_2$ if
\begin{enumerate}
\item $\op_1$ happens-before $\op_2$ in $\hist$, \label{newedgehb}
\item 
  the \methodevent{} corresponding to $\op_1$ in $\useq_2$ is before the one 
  corresponding to $\op_2$.
\end{enumerate}

Assume there is a cycle in $G$, meaning there exists $\op_1,\op_2$ such that 
$\op_1$ happens-before $\op_2$ in $\hist$, but the corresponding method events 
are in the opposite order in $\useq_2$.
\begin{itemize}
\item 
  If $\op_1,\op_2 \in L$, or $\op_1,\op_2 \in R$, this contradicts 
  $\linable{\hist'}{\useq'}$.
\item 
  If $\op_1 \in R$ and $\op_2 \in L$, this contradicts
  $\linable{\hist}{\useq}$.
\item 
  If $\op_1 \in R$ and $\op_2 = \theempty$, or if 
  $\op_1 = \theempty$ and $\op_2 \in L$, this contradicts
  $\linable{\hist}{\useq}$.
\end{itemize}
This shows that $\linable{\hist}{\useq_2}$.
Thus, we have $\linable{\hist}{\queue}$ and concludes
the proof that the \queue{} is \sbs.
\end{proof}

\begin{lemma}
\stack{} is \sbs.
\end{lemma}

\begin{proof}
\newcommand{\thepush}{a}
\newcommand{\thepop}{b}
Let $\hist$ be a \distinguished{} history, and $\useq$ a \dataword{} 
such that $\linable{\hist}{\useq}$. We have three cases to consider:

1) (very similar to case 3 of the \queue) 
$\useq$ matches $\rpushpop$ with witness $\fresh$: let $\thepush$ and 
$\thepop$ be 
respectively the \push{} and \pop{} operations corresponding to the witness.
Let $\hist' = \removefrom{\hist}{\fresh}$ and assume 
$\linable{\hist'}{\dw{\rpushpop}}$.
Let $L$ be the set of operations which are before $\thepop$ in $\useq$, and $R$ 
the ones which are after.
Let $\dats_L$ be the data-values appearing in $L$ and $\dats_R$ be the
data-values appearing in $R$. Since $\useq$ matches $\rpushpop$, we know that 
$L$ contains no unmatched \push{} operations.

Let $\useq' \in \dw{\rpushpop}$ such that 
$\linable{\hist'}{\useq'}$.
Let $\useq'_L = \project{\useq'}{\dats_L}$ and 
$\useq'_R = \project{\useq'}{\dats_R}$. 
Since $\dw{\rpushpop}$ is \closedbyerasure{}, 
$\useq'_L,\useq'_R \in \dw{\rpushpop}$. Let  
$\useq_2 = \thepush \cc \useq'_L \cc \thepop \cc \useq'_R$. 
We can show that $\useq_2 \in \dw{\rpushpop}$ by using
the derivations of $\useq'_L$ and $\useq'_R$. 

Moreover, we have $\linable{\hist}{\useq_2}$, because if the total order of 
$\useq_2$ didn't respect the \happenedbefore{} relation of $\useq_2$, it 
could only be because of four reasons, all leading to a contradiction:
\begin{itemize}
\item 
  the violation is between two $L$ operations or two $R$ operations, 
  contradicting $\linable{\hist'}{\useq'}$
\item 
  the violation is between a $L$ and an $R$ operation, 
  contradicting  $\linable{\hist}{\useq}$
\item 
  the violation is between $\thepop$ and another operation, 
  contradicting $\linable{\hist}{\useq}$
\item
  the violation is between $\thepush$ and another operation 
  contradicting $\linable{\hist}{\useq}$
\end{itemize}
This shows that $\linable{\hist}{\dw{\rpushpop}}$ and concludes case 1.

2) $\useq$ matches $\rpush$ with witness $\fresh$:
similar to case 1

3) $\useq$ matches $\rpopempty$ with witness $\fresh$: identical to case 3
of the \queue
\end{proof}
\end{proof}

\begin{lemma}
\register{} is \sbs.
\end{lemma}

\begin{proof}
Let $\hist$ be a \distinguished{} history, and $\useq$ a \dataword{} 
such that $\linable{\hist}{\useq}$ and such that $\useq$ matches the rule 
$\rreg$ with witness $\fresh$. Let $\thewrite$ and 
$\theread_1,\dots,\theread_\nreads$ be respectively the $\writevar$ and 
$\readvar$'s operations of $\hist$ corresponding to the witness.

Let $\hist' = \removefrom{\hist}{\fresh}$ and assume 
$\linable{\hist'}{\dw{\rreg}}$. Let $\useq' \in \dw{\rreg}$
such that $\linable{\hist'}{\useq'}$. 
Let 
$\useq_2 = \thewrite \cc \theread_1 \cc \theread_2 \cdots \theread_\nreads 
\cc \useq'$. By using \ruler{} $\rreg$ on $\useq'$, 
we have $\useq_2 \in \dw{\rreg}$.
Moreover, we prove that $\linable{\hist}{\useq_2}$ by contradiction. Assume
that the total order imposed by $\useq_2$ doesn't respect the \happenedbefore{}
relation of $\hist$. All three cases are not possible:
\begin{itemize}
\item 
  the violation is between two $\useq'$ operations, 
  contradicting $\linable{\hist'}{\useq'}$,
\item 
  the violation is between $\thewrite$ and another operation, \ie there is an
  operation $\op$ which happens-before $\thewrite$ in $\hist$, contradicting 
  $\linable{\hist}{\useq}$,
\item
  the violation is between some $\theread_i$ and a $\useq'$ operation, \ie there
  is an operation $\op$ which happens'before $\theread_i$ in $\hist$, 
  contradicting $\linable{\hist}{\useq}$.
\end{itemize}

Thus, we have $\linable{\hist}{\useq_2}$ and $\linable{\hist}{\dw{\rreg}}$,
which ends the proof.
\end{proof}

\begin{lemma}
\mutex{} is \sbs.
\end{lemma}

\begin{proof}
Identical to the \register{} proof, expect there is only one \unlock{} 
operation ($\theread$), instead of several \readvar{} operations
($\theread_1,\dots,\theread_\nreads$).
\end{proof}



\subsection{Regularity}

\regularity*

\begin{proof}

We have
a generic schema to build the automaton, which is first to characterize a 
violation by the existence of a cycle of some kind, and then build an automaton
recognizing such cycles. For some of the \ruler{s}, we prove that these 
cycles can always be bounded, thanks to a \emph{small model property}. For the 
others, even though the cycles can be unbounded, we can still build an automaton

(\queue) The empty automaton proves that $\rstart$ and $\renq$ are regular,
as there is no execution $\exec'$ such that $\getrule(\exec') = \rul$ and
$\notlinable{\exec'}{\matchset{\rul}}$ for $\rul \in \set{\rstart,\renq}$.
The proofs for $\renqdeq$ and $\rempty$ are more complicated and can be
found respectively in \lem{lem:renqdeq-tract} and \lem{lem:rempty-tract}

(\stack) The proofs can be found in Appendix~\ref{app:stack-tract}.

(\register{} and \mutex{}) Similarly to the \ruler{} $\renqdeq$, we can reprove
\lem{lem:smp} (with sublemmas~\ref{lem:compatible}, \ref{lem:generalization}
and \ref{lem:cycles}) to get a small model property, and build an automaton for 
the small violations.
\end{proof}

\subsection{Regularity of $\renqdeq$}

\begin{lemma}
\label{lem:smp}
Given a history $\hist$,
if $\forall \dat_1,\dat_2 \in \domof{\hist}$,
$\linable{\project{\hist}{\set{\dat_1,\dat_2}}}{\renqdeq}$,
then $\linable{\hist}{\renqdeq}$.
\end{lemma}

\begin{proof}

We first identify constraints which are sufficient to prove that 
$\linable{\hist}{\renqdeq}$.
\begin{lemma}
\label{lem:compatible}
Let $\hist$ be a history and $\fresh$ a  data value of $\domof{\hist}$. If
$\dataletter{\enqueue}{\fresh} \notafter \dataletter{\dequeue}{\fresh}$,
and for all operations $\op$, we have 
$\dataletter{\enqueue}{\fresh} \notafter \op$,
and for all \dequeue{} operations $\op$, we have
$\dataletter{\dequeue}{\fresh} \notafter \op$,
then $\hist$ is linearizable with respect to $\matchset{\renqdeq}$
\end{lemma}

\begin{proof}
We define a graph $G$ whose nodes are the element of $\hist$, and whose edges
include both the \happenedbefore{} relation as well as the constraints depicted
given by the Lemma. $G$ is acyclic by assumption and any total order 
compatible with $G$ corresponds to a linearization of $\hist$ which is in
$\matchset{\renqdeq}$.
\end{proof}

Given $\dat_1,\dat_2 \in \domof{\hist}$, we denote by
$\dat_1 \winningp{\hist}{\matchset{\rul}} \dat_2$ the fact that 
$\project{\hist}{\set{\dat_1,\dat_2}}$ is linearizable with respect to 
$\rul$,
by using $\dat_1$ as a witness for the existentially quantified $\fresh$ 
variable. We reduce the notation to $\dat_1 \winning \dat_2$ when the context 
is not ambigious.

First, we show that if the same data value can be used as a witness for 
$\fresh$ for all projections of size $2$, then we can linearize the whole 
history (using this same data value as a witness).

\begin{lemma}
\label{lem:generalization}
For $\dat_1 \in \domof{\hist}$, 
if $\forall \dat \neq \dat_1$, $\dat_1 \winning \dat$,
then $\linable{\hist}{\matchset{\renqdeq}}$.
\end{lemma}
\begin{proof}
Since $\forall \dat \neq \dat_1$, $\dat_1 \winning \dat$, the 
\happenedbefore{} relation of $\hist$ respects the constraints given by 
$\lem{lem:compatible}$, and we can conclude that 
$\linable{\hist}{\matchset{\renqdeq}}$.
\end{proof}

Next, we show the key characterization, which enables us to reduce 
non-linearizability with respect to $\matchset{\renqdeq}$ to the existence of a 
cycle in the $\notwinning$ relation.

\begin{lemma}
\label{lem:cycles}
If $\notlinable{\hist}{\matchset{\renqdeq}}$, then 
$\hist$ has a cycle
$\dat_1 \notwinning \dat_2 \notwinning \dots \notwinning \dat_\ndats
\notwinning \dat_1$
\end{lemma}

\begin{proof}
Let $\dat_1 \in \domof{\hist}$.
By \lem{lem:generalization}, we know there exists $\dat_2 \in \domof{\hist}$ 
such that $\dat_1 \notwinning \dat_2$. 
Likewise, we know there exists $\dat_3 \in \domof{\hist}$ such that
$\dat_2 \notwinning \dat_3$. We continue this construction until we form a 
cycle.
\end{proof}

We can now prove the small model property.
Assume $\notlinable{\hist}{\rul}$. By \lem{lem:cycles}, it has a cycle
$\dat_1 \notwinning \dat_2 \notwinning \dots \notwinning \dat_\ndats
\notwinning \dat_1$. If there exists a data-value $\fresh$ such that 
$\dataletter{\dequeue}{\fresh}$ happens-before $\dataletter{\enqueue}{\fresh}$, 
then $\notlinable{\project{\hist}{\set{\fresh}}}{\renqdeq}$, which contradicts
our assumptions.

For each $i$, there are two possible reasons for which
$\dat_i \notwinning \dat_{(\modulo{i}{\ndats}) + 1}$. The first one is that
$\dataletter{\enqueue}{\dat_i}$ is not minimal in the subhistory of size $2$ 
(reason (a)).
The second one is that $\dequeue_{\dat_i}$ is not minimal with respect 
to the $\dequeue$ operations (reason (b)).

We label each edge of our cycle by either (a) or (b), depending on which one
is true (if both are true, pick arbitrarily). 
Then, using the interval order property, we have that, if
$\dat_i \notwinning \dat_{(\modulo{i}{\ndats}) + 1}$ for reason (a), and
$\dat_j \notwinning \dat_{(\modulo{j}{\ndats}) + 1}$ for reason (a) as well,
then either $\dat_i \notwinning \dat_{(\modulo{j}{\ndats}) + 1}$, or
$\dat_j \notwinning \dat_{(\modulo{i}{\ndats}) + 1}$ (for reason (a)). This enables
us to reduce the cycle and leave only one edge for reason (a).

We show the same property for (b). This allows us to reduce the cycle to a
cycle of size $2$ (one edge for reason (a), one edge for reason (b)). If 
$\dat_1$ and $\dat_2$ are the two data-values appearing in the cycle, we have:
$\notlinable{\project{\hist}{\set{\dat_1,\dat_2}}}{\renqdeq}$, which is a 
contradiction as well.
\end{proof}

\begin{lemma}
\label{lem:renqdeq-tract}
The \ruler{} $\renqdeq$ is \tractable.
\end{lemma}
\begin{proof}
We prove in \lem{lem:smp} that a \distinguished{} 
history $\hist$ has a projection $\hist'$ such that 
$\getrule(\hist') = \renqdeq$ and 
$\notlinable{\hist'}{\matchset{\renqdeq}}$
if and only if it has such a projection on $1$ or $2$ data-values.
Violations of histories with two values are:
$i)$ there is a value $\fresh$ such that $\dataletter{\dequeue}{\fresh}$
happens-before $\dataletter{\enqueue}{\fresh}$ 
(or $\dataletter{\enqueue}{\fresh}$ doesn't exist in the history) or
$ii)$ there are two operations $\dataletter{\dequeue}{\fresh}$ in $\hist$ or,
$iii)$ there are two values $\fresh$ and $y$ such that 
$\dataletter{\enqueue}{\fresh}$ happens-before $\dataletter{\enqueue}{y}$, and
$\dataletter{\dequeue}{y}$ happens-before $\dataletter{\dequeue}{\fresh}$ 
($\dataletter{\dequeue}{\fresh}$ doesn't exist in the history).

The automaton $\autof{\renqdeq}$ in \fig{fig:queue-enqdeq} recognizes all such 
small violations (top branch for $i$, middle branch for $ii$, bottom branch
for $iii$).

Let $\impl$ be any data-independent implementation. We show that
\[
  \autof{\renqdeq} \cap \impl \neq \emptyset \iff
  \exists \exec \in \restrict{\impl}, \exec' \in \projections{\exec}.\ 
    \getrule(\exec') = \renqdeq \land \notlinable{\exec'}{\matchset{\renqdeq}}
\]

$(\Rightarrow)$ Let $\exec \in \impl$ be an execution which is accepted
by $\autof{\renqdeq}$. By data independence, let $\restrict{\exec} \in \impl$ 
$r$ a renaming such that $\exec = r(\restrict{\exec})$, and assume without 
loss of generality that $r$ doesn't rename the data-values $1$ and $2$. 
If $\exec$ is accepted by the top or middle branch of $\autof{\renqdeq}$, we 
can project $\restrict{\exec}$ on value $1$ to obtain a projection $\exec'$ such
that $\getrule(\exec') = \renqdeq$ and 
$\notlinable{\exec'}{\matchset{\renqdeq}}$.
Likewise, if $\exec$ is accepted by the bottom branch, we can project 
$\restrict{\exec}$ on $\set{1,2}$, and obtain again a projection $\exec'$ such 
that $\getrule(\exec') = \renqdeq$ and 
$\notlinable{\exec'}{\matchset{\renqdeq}}$.

$(\Leftarrow)$
Let $\restrict{\exec} \in \restrict{\impl}$ such that there is a projection 
$\exec'$ such that $\getrule(\exec') = \renqdeq$ and 
$\notlinable{\exec'}{\matchset{\renqdeq}}$. As recalled at the beginning of 
the proof, we know $\restrict{\exec}$ has to contain a violation of type $i$, 
$ii$, or $ii$.
If it is of type $i$ or $ii$, we define the renaming $r$, which maps $\fresh$ 
to $1$, and all other data-values to $2$. The execution $r(\restrict{\exec})$ 
can then be recognized by the top or middle branch of $\autof{\renqdeq}$ and 
belongs to $\impl$ by data independence. 

Likewise, if it is of type $iii$, $r$ will map $\fresh$ to 
$1$, and $y$ to $2$, and all other data-values to $3$, so that 
$r(\restrict{\exec})$ can be recognized by the bottom branch of 
$\autof{\renqdeq}$.
\end{proof}

\subsection{Regularity of $\rempty$}

We first define the notion of \emph{\gap}, which intuitively corresponds to a
point in an execution where the \queue{} could be empty.

\begin{definition}
Let $\hist$ be a \distinguished{} history and $\op$ an operation of $\hist$. 
We say that $\hist$ has a \emph{\gap{} on operation $\op$} if there is a 
partition of the operations of $\hist$ into $L \uplus R$ satisfying:
\begin{itemize}
\item $L$ has no unmatched $\enqueue$ operation, and
\item no operation of $R$ happens-before an operation of $L$ or $\op$, and
\item no operation of $L$ happens-after $\op$.
\end{itemize}
\end{definition}

\begin{lemma}
\label{lem:gap}
A \distinguished{} history $\hist$ has a projection $\hist'$ such that 
$\getrule(\hist') = \rempty$ and $\notlinable{\hist'}{\matchset{\rempty}}$ if
and only there exists a $\deqempty$ operation $\theempty$ in $\hist$ such that 
there is no gap on $\theempty$.
\end{lemma}
\begin{proof}
$(\Rightarrow)$ Assume there exists a projection $\hist'$ such that 
$\getrule(\hist') = \rempty$ and $\notlinable{\hist'}{\matchset{\rempty}}$. 
Let $\op$ be a $\deqempty$ operation in $\hist'$ (exists by definition of 
$\getrule$).

Assume by contradiction that there is a gap on $\theempty$. 
By the properties of the gap, we can linearize $\hist'$ into a \dataword{} 
$\useq \cc \theempty \cc \vseq$ where $\useq$ and $\vseq$ respectively contain 
the $L$ and $R$ operations of the partition.

$(\Leftarrow)$ Assume there exists a $\deqempty$ operation $\theempty$ in 
$\hist$ such that there is no gap on $\theempty$. Let $\hist'$ be the projection 
which contains all the operations of $\hist$ as well as $\theempty$, except the 
other $\deqempty$ operations.

Assume by contradiction that there exists a \dataword{} 
$\wseq \in \matchset{\rempty}$ such that $\linable{\hist'}{\wseq}$. By
definition of $\matchset{\rempty}$, $\wseq$ can be decomposed into 
$\useq \cc \theempty \cc \vseq$ such that $\useq$ has no unmatched 
operation. Let $L$ be the operations of $\useq$, and $R$ 
the operation of $\vseq$. Since $\linable{\hist'}{\wseq}$, the partition 
$L \uplus R$ forms a gap on operation $\theempty$.
\end{proof}

We exploit the characterization of \lem{lem:gap} by showing how we can
recognize the existence of \gap{s} in the next two lemmas. First, we define
the notion of \emph{\leftrightconstraint{s}} of an operation, and show that
this constraints have a solution if and only if there is a gap on the 
operation.

\begin{definition}
Let $\hist$ be a distinguished history, and $\op$ an operation of $\hist$.
The \leftrightconstraint{s} of $\op$ is the graph $G$ where:
\begin{itemize}
\item 
  the nodes are $\domof{\hist}$, the data-values of $\hist$, to which we
  add a node for $\op$,
\item 
  there is an edge from data-value $\dat_1$ to $\op$ if 
  $\dataletter{\enqueue}{\dat_1}$ happens-before $\op$,
\item
  there is an edge from $\op$ to data-value $\dat_1$ if 
  $\op$ happens-before $\dataletter{\dequeue}{\dat_1}$,
\item
  there is an edge from data-value $\dat_1$ to $\dat_2$ if 
  $\dataletter{\enqueue}{\dat_1}$ happens before 
  $\dataletter{\dequeue}{\dat_2}$.
\end{itemize}
\end{definition}

\begin{lemma}
\label{lem:coloring}
Let $\hist$ be a \distinguished{} history and $\op$ an operation of $\hist$.
Let $G$ be the graph representing the \leftrightconstraint{s} of $\op$.
There is a \gap{} on $\op$ if and only if $G$ has no cycle going through
$\op$.

\begin{proof}
$(\Rightarrow)$ Assume that there is a \gap{} on $\op$, and let $L \uplus R$ 
be a partition corresponding to the \gap{}. Assume by contradiction there is 
a cycle 
$\dat_\ndats \rightarrow \dots \rightarrow \dat_1 \rightarrow \op \rightarrow 
\dat_\ndats$ in $G$ (which goes through $\op$).
By definition of $G$, and since $\op \rightarrow \dat_\ndats$, and by definition
of a \gap{}, we know that all operations with data-value $\dat_\ndats$ must be 
in $R$. Since $\dat_\ndats \rightarrow \dat_{\ndats-1}$, the operations with 
data-value $\dat_{\ndats-1}$ must be in $R$ as well. We iterate this reasoning 
until we deduce that $\dat_1$ must be in $R$, contradicting the fact that 
$\dat_1 \rightarrow \op$.

$(\Leftarrow)$ Assume there is no cycle in $G$ going through $\op$. Let $L$ 
be the set of operations having a data-value $\dat$ which has a path to $\op$ 
in $G$, and let $R$ be the set of other operations. By definition of the 
\leftrightconstraint{s} $G$, the partition $L \uplus R$ forms a \gap{} for 
operation $\op$.
\end{proof}
\end{lemma}

\begin{corollary}
\label{coro:cover}
A \distinguished{} history $\hist$ has a projection $\hist'$ such that 
$\getrule(\hist') = \rempty$ and $\notlinable{\hist'}{\matchset{\rempty}}$ 
if and only if it has a $\deqempty$ operation $\theempty$ and data-values 
$\dat_1,\dots,\dat_\ndats \in \domof{\hist}$ such that:
\begin{itemize}
\item 
  $\dataletter{\enqueue}{\dat_1}$ happens-before $\theempty$ in $\hist$, and
\item 
  $\dataletter{\enqueue}{\dat_i}$ happens before 
  $\dataletter{\dequeue}{\dat_{i-1}}$ in $\hist$ for $i > 1$, and
\item
  $\theempty$ happens-before $\dataletter{\dequeue}{\dat_\ndats}$, or 
  $\dataletter{\dequeue}{\dat_\ndats}$ doesn't exist in $\hist$.
\end{itemize}
We say that $\theempty$ is \emph{covered} by $\dat_1,\dots,\dat_\ndats$.
\end{corollary}

\begin{proof}
By definition of the \leftrightconstraint{s}, and following from 
Lemmas~\ref{lem:gap} and \ref{lem:coloring}.
\end{proof}

\begin{lemma}
\label{lem:rempty-tract}
The \ruler{} $\rempty$ is \tractable.
\end{lemma}

\begin{proof}
See \sect{sec:reduction}.
\end{proof}

\subsection{Regularity of the \stack{} \ruler{s}}
\label{app:stack-tract}

\begin{lemma}
\label{lem:gaps}
A \distinguished{} history $\hist$ has a projection $\hist'$ such that 
$\getrule(\hist') = \rpushpop$ and 
$\notlinable{\hist'}{\matchset{\rpushpop}}$ if and only if there exists 
a projection such that $\getrule(\hist') = \rpushpop$ and either
\begin{itemize}
\item 
  there exists an unmatched $\dataletter{\pop}{\dat}$ operation in $\hist'$, 
  or
\item 
  there is a $\dataletter{\pop}{\dat}$ which happens-before
  $\dataletter{\push}{\dat}$ in $\hist'$, or
\item for all
  $\dataletter{\push}{\dat}$ operations minimal in $\hist'$,
  there is no \gap{} on 
  $\dataletter{\pop}{\dat}$ in $\removefrom{\hist'}{\dat}$.
\end{itemize}
\end{lemma}

\begin{proof}
Similar to \lem{lem:gap}.
\end{proof}

\begin{lemma}
\label{lem:stackcover}
A \distinguished{} history $\hist$ has a projection $\hist'$ such that 
$\getrule(\hist') = \rpushpop$ and 
$\notlinable{\hist'}{\matchset{\rpushpop}}$ if and only if either:
\begin{itemize}
\item there exists an unmatched $\dataletter{\pop}{\dat}$ operation, or
\item there is a $\dataletter{\pop}{\dat}$ which happens-before
$\dataletter{\push}{\dat}$, or
\item 
there exist
a data-value $\dat \in \domof{\hist}$ and data-values 
$\dat_1,\dots,\dat_\ndats \in \domof{\hist}$ such that
\begin{itemize}
\item 
  $\dataletter{\push}{\dat}$ happens-before $\dataletter{\push}{\dat_i}$
  for every $i$,
\item
  $\dataletter{\pop}{\dat}$ is covered by $\dat_1,\dots,\dat_\ndats$.
\end{itemize}
\end{itemize}
\end{lemma}

\begin{proof}
$(\Leftarrow)$ We have three cases to consider
\begin{itemize}
\item there exists an unmatched $\dataletter{\pop}{\dat}$ operation:
define $\hist' = \project{\hist}{\set{\dat}}$,
\item there is a $\dataletter{\pop}{\dat}$ which happens-before 
$\dataletter{\push}{\dat}$:
define $\hist' = \project{\hist}{\set{\dat}}$,
\item 
there exist
a data-value $\dat \in \domof{\hist}$ and data-values 
$\dat_1,\dots,\dat_\ndats \in \domof{\hist}$ such that
\begin{itemize}
\item 
  $\dataletter{\push}{\dat}$ happens-before $\dataletter{\push}{\dat_i}$
  for every $i$
\item
  $\dataletter{\pop}{\dat}$ is covered by $\dat_1,\dots,\dat_\ndats$.
\end{itemize}

Define $\hist' = \project{\hist}{\set{\dat,\dat_1,\dots,\dat_\ndats}}$. We 
have $\getrule(\hist') = \rpushpop$ because $\hist'$ doesn't contain 
$\popempty$ operations nor unmatched \push{} operations. Assume 
by contradiction that
$\linable{\hist'}{\matchset{\rpushpop}}$, and let 
$\wseq \in \matchset{\rpushpop}$ such that $\linable{\hist'}{\useq}$.
Since
$\dataletter{\push}{\dat}$ happens-before $\dataletter{\push}{\dat_i}$
  (for every $i$)
the witness $\fresh$ of $\wseq \in \matchset{\rpushpop}$ has to be the 
data-value $\dat$. This means that 
$\wseq = \dataletter{\push}{\dat}\cc \useq \cc 
\dataletter{\pop}{\dat} \cc \vseq$ for some $\useq$ and $\vseq$ with no 
unmatched $\push$.

Thus, there is a \gap{} on operation 
$\dataletter{\pop}{\dat}$ in $\removefrom{\hist'}{\dat}$, and that
$\dataletter{\pop}{\dat}$ cannot be covered by $\dat_1,\dots,\dat_\ndats$.

\end{itemize}

$(\Rightarrow)$ Let $\hist'$ be a projection of $\hist$ such that 
$\getrule(\hist') = \rpushpop$ and
$\notlinable{\hist'}{\matchset{\rpushpop}}$. Assume there are no 
unmatched $\dataletter{\pop}{\dat}$ operation, and that for every $\dat$,
$\dataletter{\pop}{\dat}$ doesn't happens-before $\dataletter{\push}{\dat}$.
This means that $\hist'$ is made of pairs of 
$\dataletter{\push}{\dat}$ and $\dataletter{\pop}{\dat}$ operations.

Let $\dataletter{\push}{\dat}$ be a \push{} \operation{} which is minimal in 
$\hist'$. We know there is one, because we assumed that 
$\getrule(\hist') = \rpushpop$, and we know that there is a \push{} which is
minimal because for every $\dat$, $\dataletter{\pop}{\dat}$ doesn't 
happens-before $\dataletter{\push}{\dat}$.

By \lem{lem:gaps}, we know that there is no \gap{} on 
$\dataletter{\pop}{\dat}$. Similarly to \lem{lem:coloring} and
\coro{coro:cover}, we deduce
that there are data-values
$\dat_1,\dots,\dat_\ndats \in \domof{\hist'}$ such that 
$\dataletter{\pop}{\dat}$ is covered by 
$\dat_1,\dots,\dat_\ndats$. Our goal is now to prove that we can choose
$\dat$ and $\dat_1,\dots,\dat_\ndats$ such that, besides these properties,
we also have that
$\dataletter{\push}{\dat}$ happens-before $\dataletter{\push}{\dat_i}$
for every $i$.
Assume there exists $i$ such that $\dataletter{\push}{\dat}$ 
doesn't happen-before $\dataletter{\push}{\dat_i}$.
We have two cases, either $\dataletter{\pop}{\dat}$ is covered by 
$\dat_1,\dots,\dat_{i-1},\dat_{i+1},\dots,\dat_\ndats$, in which case
we can just get rid of $\dat_i$; or this is not the case, and we can 
choose our new $\dat$ to be $\dat_i$ and remove $\dat_i$ from the list
of data-values. We iterate this until we have a data-value 
$\dat \in \domof{\hist}$ such that
\begin{itemize}
\item 
  $\dataletter{\push}{\dat}$ happens-before $\dataletter{\push}{\dat_i}$
  for every $i$,
\item
  $\dataletter{\pop}{\dat}$ is covered by $\dat_1,\dots,\dat_\ndats$.
\end{itemize}
\end{proof}

\begin{lemma}
The \ruler{} $\rpushpop$ is \tractable.
\end{lemma}

\begin{proof}
The automaton \fig{fig:stack-pushpop} recognizes the violations given by
\lem{lem:stackcover}. 
The proof is then similar to \lem{lem:rempty-tract}.
\end{proof}

\begin{lemma}
The \ruler{} $\rpush$ is \tractable.
\end{lemma}

\begin{proof}
We can make a characterization of the violations similar to 
\lem{lem:stackcover}. This \ruler{} is in a way simpler, because the $\push$
in this \ruler{} plays the role of the $\pop$ in $\rpushpop$.
\end{proof}

\begin{lemma}
The \ruler{} $\rpopempty$ is \tractable.
\end{lemma}

\begin{proof}
Identical to \lem{lem:rempty-tract} (replace $\enqueue$ by $\push$,
$\dequeue$ by $\pop$, and $\deqempty$ by $\popempty$).
\end{proof}

%
%
%
%
%
%


\subsection{Regular automata used to prove regularity}

\begin{figure}[hbt]
\begin{center}
\begin{tikzpicture}[x=1.8cm, y=1.5cm, ->]

\node[state, initial, initial text=] (a) {$q_0$};
\node[state,right=1 of a] (ab) {$q_1$};
\node[state,right=1 of ab] (b) { $q_1$ };
\node[state,right=1 of b] (c) { $q_2$ };
\node[state,right=1 of c,accepting] (d) { $q_3$ };

\path (a) edge [loop above] node[above] {$\thealphabet{3}$} (a);
\path (a) edge [loop below] node[below] {$\callevent{\dequeue}{2}{}$} (a);
\path (ab) edge [loop above] node[above] {$\thealphabet{3}$} (ab);
\path (b) edge [loop above] node[above] {$\thealphabet{3}$} (b);
\path (c) edge [loop above] node[above] {$\thealphabet{3}$} (c);
\path (d) edge [loop above] node[above] {$\thealphabet{3}$} (d);

\path (a) edge node[above] {$\callevent{\enqueue}{1}{}$} (ab);
\path (ab) edge node[above] {$\retevent{\enqueue}{1}{}$} (b);
\path (b) edge node[above] {$\callevent{\enqueue}{2}{}$} (c);
\path (c) edge node[above] {$\retevent{\dequeue}{2}{}$} (d);

\node[state, initial, initial text=,above=1.6 of a] (i) {$q_4$};
\node[state,right=1 of i] (j) {$q_5$};
\node[state,right=1 of j, accepting] (k) { $q_6$ };
\path (i) edge [loop above] node[above] {$\thealphabet{1},\thealphabet{2}$} (i);
\path (j) edge [loop above] node[above] {$\thealphabet{1},\thealphabet{2}$} (j);
\path (k) edge [loop above] node[above] {$\thealphabet{1},\thealphabet{2}$} (k);
\path (i) edge [loop below] node[below] {$\callevent{\dequeue}{1}{}$} (i);
\path (i) edge node[above] {$\retevent{\dequeue}{1}{}$} (j);
\path (j) edge node[above] {$\retevent{\dequeue}{1}{}$} (k);

\node[state, initial, initial text=,above=3.8 of a] (p) {$q_7$};
\node[state,right=1 of p, accepting] (q) {$q_8$};
\path (p) edge [loop above] node[above] {$\thealphabet{2}$} (p);
\path (q) edge [loop above] node[above] {$\thealphabet{2}$} (q);
\path (p) edge [loop below] node[below] {$\callevent{\dequeue}{1}{}$} (p);
\path (p) edge node[above] {$\retevent{\dequeue}{1}{}$} (q);

\end{tikzpicture}
\end{center}
\caption{A non-deterministic automaton recognizing $\renqdeq$ violations.
The top branch recognizes executions which have a \dequeue{} with no 
corresponding \enqueue{}. The middle branch recognizes two \dequeue{}'s
returning the same value, which is not supposed to happen in a 
\distinguished{} execution. The bottom branch recognizes FIFO violations.
By the closure properties of \implementation{s}, we can assume 
the $\callevent{\dequeue}{2}{}$ are at the beginning.}
\label{fig:queue-enqdeq}
\end{figure}

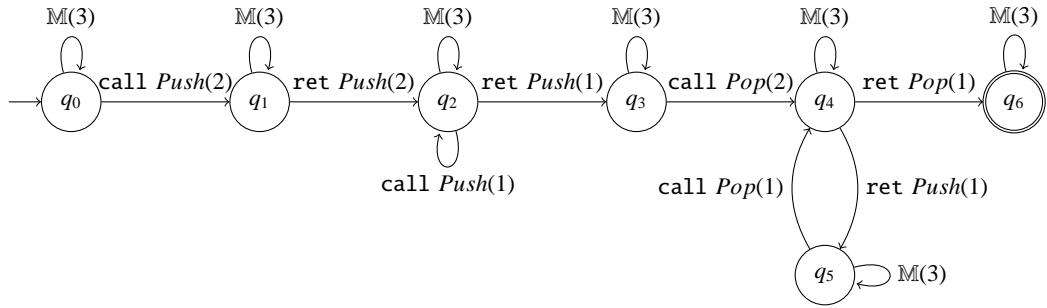
\begin{figure}[hbt]
\begin{center}
\begin{tikzpicture}[x=1.7cm, y=1.5cm, ->]

\footnotesize

\node[state, initial, initial text=] (j) {$q_0$};
\node[state,right=1 of j] (l) {$q_1$};
\node[state,right=1 of l] (a) {$q_2$};
\node[state,right=1 of a] (b) { $q_3$ };
\node[state,right=1 of b] (c) { $q_4$ };
\node[state,below=1 of c] (d) { $q_5$ };
\node[state, accepting, right=1 of c] (e) { $q_6$ };

\path (j) edge [loop above] node[above] {$\thealphabet{3}$} (j);
\path (l) edge [loop above] node[above] {$\thealphabet{3}$} (l);
\path (a) edge [loop above] node[above] {$\thealphabet{3}$} (a);
\path (b) edge [loop above] node[above] {$\thealphabet{3}$} (b);
\path (c) edge [loop above] node[above] {$\thealphabet{3}$} (c);
\path (d) edge [loop right] node[right] {$\thealphabet{3}$} (d);
\path (e) edge [loop above] node[above] {$\thealphabet{3}$} (e);

\path (a) edge [loop below] node[below] {$\callevent{\push}{1}{}$} (a);
\path (j) edge node[above] { $\callevent{\push}{2}{}$ } (l);
\path (l) edge node[above] { $\retevent{\push}{2}{}$ } (a);
\path (a) edge node[above] { $\retevent{\push}{1}{}$ } (b);
\path (b) edge node[above] { $\callevent{\pop}{2}{}$ } (c);
\path (c) edge node[above] { $\retevent{\pop}{1}{}$ } (e);
\path (c) edge [bend left] node[right] { $\retevent{\push}{1}{}$ } (d);
\path (d) edge [bend left] node[left] { $\callevent{\pop}{1}{}$ }  (c);
\end{tikzpicture}
\end{center}
\caption{An automaton recognizing $\rpushpop$ violations. Here we have a
$\push(2)$ operation, whose corresponding $\pop(2)$ operation is covered
by $\push(1)/\pop(1)$ pairs. The $\push(2)$ happens-before all the pairs.
Intuitively, the element $2$ cannot be popped from the \stack{} there is 
always at least an element $1$ above it in the \stack{} (regardless of how
linearize the execution).}
\label{fig:stack-pushpop}
\end{figure}

  }{}

\end{document}